# The Effects of Air-Conditioning and Road-Traffic Noise on Perceived, Cognitive, and EEG Responses in a University Classroom

Yuanzhi Su[1], Cynthia (Huiying) Hou[1, *]

[1]Department of Building Environment and Energy Engineering, The Hong Kong Polytechnic University, Hong Kong

## Abstract

Air-conditioning and road-traffic noise are common in classrooms, and their effects are often judged by cognitive performance. However, a sound that leaves performance unchanged may still be perceived differently or alter brain activity. In a within-participant design, this study tested whether perceived, cognitive, and EEG responses are consistent, and whether perceived and EEG responses distinguish conditions that performance does not. Sixteen students completed cognitive tasks in a university classroom under air-conditioning and road-traffic noise, each played back at 55, 60, and 65 dBA. Participants rated acoustic appraisal and perceived workload after each condition, and EEG was recorded during the tasks. The results show that cognitive performance did not differ significantly between conditions in any task. By contrast, noise was rated louder, less pleasant, and more arousing than a no-noise control, and all three ratings varied with sound level; perceived workload was also higher under noise. Task-period EEG power also showed significant differences between sources or levels in specific tasks and scalp regions. Within participants, relative EEG band power covaried mainly with loudness and pleasantness ($|r_{rm}|$ = .27–.55); louder and less pleasant ratings coincided with higher relative theta and lower relative gamma power. Associations with arousal and performance were rare, and none were found for workload. The three response types were thus only partly consistent, and perceived responses and EEG distinguished conditions that performance did not. Non-significant performance differences do not mean the conditions were equivalent for students. Assessments of air-conditioning and road-traffic noise in classrooms should therefore consider perceived responses alongside cognitive performance, with EEG providing complementary information on neural activity during tasks.

## 1. Introduction

Students spend a large share of their day in classrooms, approximately 45% of their daily time on average (PISA, 2019) . Much of that time goes to listening to instruction, reading, and solving problems. These activities place sustained demands on attention and memory, so the classroom environment matters for learning as well as for comfort. Ventilation shows this clearly: adequate ventilation improves indoor air quality and has been associated with better student performance (Wargocki et al., 2020). The way a classroom is ventilated and cooled, however, also shapes what students hear (Pellegatti et al., 2023) . When a classroom relies on air-conditioning with the windows closed, the air-conditioning system becomes a continuous indoor sound source. When the windows are opened for natural ventilation, outdoor sounds such as road traffic enter the room (Pellegatti et al., 2023, 2026). Exposure to air-conditioning or road-traffic sound is therefore closely tied to how the classroom itself is operated. The two sounds also differ markedly in character. Sound from mechanical ventilation like air-conditioning systems is largely stationary and tends to be perceived as monotonous. Road-traffic sound, by contrast, is marked by the passage of individual vehicles and tends to be perceived as eventful and chaotic (Pellegatti et al., 2026). The effect of a sound on cognition depends on its spectral and temporal characteristics as well as on its level (Klatte et al., 2013; Pellegatti et al., 2023). The two sounds may therefore affect students differently even when presented at the same level. Their practical relevance and their acoustic contrast make them a useful pair for examining how background sound in classrooms affects the people exposed to it.

* Corresponding author: Cynthia (Huiying) Hou. Email: cynthia.hou@polyu.edu.hk

Previous research shows that background sound in learning environments can have both negative and positive effects. Traffic noise has generally been associated with poorer performance, greater effort, and more annoyance. Mechanical ventilation sound tends to increase annoyance, whereas natural sounds tend to reduce it (Pellegatti et al., 2023) . The size and even the direction of these effects, however, vary across tasks, outcome measures, populations, and experimental settings (Chen & Ou, 2021; Meng et al., 2021; Pellegatti et al., 2026; Yang & Jeon, 2023; L. Zhang & Ma, 2022) . Changes in sound level, for example, have been associated with differences in listening, memory, and attention outcomes, as well as in soundscape evaluation. Yet comparable effects are not consistently observed across tasks or sound sources.

Part of this inconsistency may lie in the outcome on which conclusions are based. When the effect of a sound is judged by cognitive performance alone, a sound that does not significantly change accuracy or response time is easily regarded as having no effect. Cognitive performance, however, is only one of the ways in which students respond to their acoustic environment. The same sound is also perceived and evaluated, it may change how demanding a task feels, and it is accompanied by neural activity while the task is carried out. These responses need not change together. At a realistic sound level, students' soundscape evaluations have been found to differ clearly between sounds that left their accuracy unchanged (Pellegatti et al., 2026) . A single outcome can therefore show that an acoustic condition has an effect, but not that it has none.

This perspective motivated the present study. A within-participant experiment was conducted in a university classroom. Recordings of air-conditioning and road-traffic noise were each played back at 55, 60, and 65 dBA, and a control condition without added noise was also included, giving seven acoustic conditions. In each condition, participants performed cognitive tasks covering attention and information processing while EEG was recorded continuously. After the tasks, they completed questionnaires on acoustic appraisal and perceived workload. Data from three response domains were thus obtained for every participant and condition: perceived responses from the questionnaires, cognitive performance (accuracy, response time, and a performance index), and task-period EEG spectral power. The contribution of this study lies not in collecting these data together, which previous studies have also done (Section 1.2), but in using them to answer two questions that separate studies cannot. The first is whether responses in the three domains are consistent with one another in the classroom context. The second is whether perceived and EEG responses can still distinguish between acoustic conditions when cognitive performance does not.

### 1.1 Cognitive performance as an indicator of acoustic effects

Cognitive performance provides the most direct evidence of whether a sound interferes with the activities through which students learn, and it is usually measured with standardised cognitive tests. The abilities assessed range from basic functions to complex abilities. Basic functions, such as attention and memory, underpin learning. Complex abilities, such as reading comprehension and mathematical calculation, build on these functions and more closely resemble classroom tasks (Caviola et al., 2021; Connolly et al., 2019; Pellegatti et al., 2023). The tests yield objective measures, typically accuracy and response time, that can be obtained repeatedly under controlled conditions and compared across sounds within the same participants. They therefore capture the immediate effect of an acoustic condition. Academic attainment, e.g., (Shield & Dockrell, 2008), by contrast, accumulates over long periods and depends on many factors besides the acoustic environment, such as standardized test in literacy.

With such tests, road-traffic and other external noise has been associated with lower accuracy in student's calculation (Ljung et al., 2009; Papanikolaou et al., 2015). Anthropogenic noise has been associated with slower responses in mathematical tasks (Caviola et al., 2021). These effects, however, were mostly obtained at relatively high sound levels, between about 55 and 80 dB(A) (Caviola et al., 2021; Connolly et al., 2019; Ljung et al., 2009; Papanikolaou et al., 2015). At levels closer to those found in real classrooms, performance effects appear weaker and less consistent. (Pellegatti et al., 2026) compared a quiet condition with natural ventilation sounds, including road traffic, and with mechanical ventilation sound. All sounds were presented at 47 dB(A) to match measured classroom conditions. The sound conditions had no effect on accuracy in calculation or on reading comprehension. The only effects concerned calculation response times, which were shorter under playground sound for difficult equations and longer under mechanical ventilation sound for easy ones.

A non-significant difference in performance does not mean, however, that two acoustic conditions were experienced or processed in the same way, because several mechanisms can keep performance stable while other responses change. People can protect their performance under adverse conditions by investing additional effort (Hockey, 2011) . They can also focus attention more narrowly on a demanding task so that background sound interferes less, a mechanism described as shielding against distraction (Halin et al., 2014). Task difficulty also matters. Easy tasks may leave enough spare capacity for a sound to have no visible effect, and very demanding ones may already use up the resources that the sound would otherwise disrupt (Caviola et al., 2021; Meinhardt‐Injac et al., 2015). Finally, sounds and sound levels differ in the arousal they evoke (Alvar & Francis, 2024; Bradley & Lang, 2000), and arousal is related to performance through an inverted U-shaped function (Yerkes & Dodson, 1908). Near the optimum of this function, a sound may change arousal without producing any measurable change in performance. (Pellegatti et al., 2026) interpreted their response-time effects in these terms. Because they measured neither physiological nor psychological indicators of arousal, however, they could not test this interpretation.

Performance alone therefore shows whether a sound disrupts a task. It does not show whether a sound that leaves performance unchanged is nonetheless perceived differently, demands more effort, or alters the neural state in which the task is performed. Answering these questions requires responses from other domains.

### 1.2 Subjective and EEG responses and their consistency with performance

Subjective responses describe how students experience a sound and the task they perform while hearing it. The soundscape approach treats the acoustic environment as it is perceived and understood by people in context (Iso, 2014). The perception of sound environments has been described along two main dimensions, pleasantness and eventfulness (Axelsson et al., 2010). Ratings of valence and arousal, in turn, capture affective reactions to individual sounds (Bradley & Lang, 2000). Such evaluations have recently been applied to classrooms (Pellegatti et al., 2026; Visentin et al., 2023), where they distinguish clearly between sounds. Mechanical ventilation sound has been rated as monotonous, road traffic as chaotic, and birdsong as pleasant (Pellegatti et al., 2026). In the present study, acoustic appraisal comprises perceived loudness, pleasantness, and arousal (Visentin et al., 2023). Perceived workload captures a different aspect of the subjective response, namely how demanding the task feels under a given sound. It is commonly assessed with the NASA Task Load Index (Hart, 2006). Because performance can be maintained through additional effort (Hockey, 2011), workload ratings can reveal a cost of the acoustic environment that performance does not show. Physiological measures add an objective and continuous perspective. Among them, EEG is particularly suited to studying cognitive activity. It records brain activity with high temporal resolution while the task is being performed, without interrupting it. Its spectral measures are linked to the processes that background sound is expected to affect. Frontal theta power tends to increase with working-memory load and mental workload, whereas alpha power tends to decrease as attentional demands rise (Borghini et al., 2014; Klimesch, 1999). EEG has accordingly been used to examine spectral, event-related, and neural-tracking responses under environmental noise and task demand (Chikhi et al., 2022; Eqlimi et al., 2020; Mohammadi et al., 2023; Pieper et al., 2021; N. Zhang et al., 2025). Compared with peripheral indicators of arousal such as skin conductance or heart rate (Alvar & Francis, 2024; Giannakakis et al., 2019), EEG reflects cortical activity more directly. When it is recorded during task performance (task-period EEG), it can therefore indicate whether a sound changes the neural state in which a task is carried out, even when performance remains unchanged.

Studies combining these response domains provide important precedents (Table 1). In classrooms and classroom-related settings, acoustic perception and cognitive performance have been examined together under traffic, ventilation-related, speech, and other sounds (Lee & Jeon, 2013; Yang & Jeon, 2023). EEG has been recorded during learning or cognitive activity under environmental noise and related to subsequent learning, attention, or subjective workload (Eqlimi et al., 2020; Mohammadi et al., 2023). Subjective evaluation and EEG have been combined across multiple sound sources and sound levels (N. Zhang et al., 2025). Subjective, cognitive, and EEG measures have also been collected within the same experiment (Chong et al., 2025; Rosenkranz et al., 2026). More broadly, physiological responses and emotion have been examined as pathways through which indoor environmental quality affects task performance (Choi et al., 2023). Multidomain measurement is therefore not new.

Collecting several types of response in the same experiment, however, does not by itself show whether they agree. Establishing consistency requires relating the responses to one another within the same participants and under the same acoustic conditions. In their classroom study, for example, (Pellegatti et al., 2026) obtained soundscape ratings and cognitive performance from the same students. They observed a tendency for more eventful sounds to accompany faster calculation, but they left a combined analysis of subjective evaluations and performance to future research. Comparisons across studies cannot close this gap. Studies differ in participants, stimuli, sound levels, tasks, and EEG measures, which makes it difficult to tell whether a discrepancy between domains reflects the responses themselves or the way each study was conducted. A second question follows from the first. If the response domains are not fully consistent, some may be more sensitive to the acoustic environment than others. As discussed in Section 1.1, performance differences between sounds may be small or absent at realistic levels, while subjective evaluations of the same sounds still differ. It remains unclear whether EEG responses also distinguish conditions that performance does not, and whether subjective and EEG responses agree in doing so. If they do, a non-significant performance result would indicate only that performance was maintained. It would not mean that the acoustic conditions were equivalent for the students exposed to them.

### 1.3 Study aims and research questions

The studies reviewed above suggest that air-conditioning and road-traffic noise can affect students through several response domains. These domains need not change together, and a condition that leaves one of them unchanged may still affect another. The present study therefore measured all three domains in the same participants across the same acoustic conditions to address three research questions:

RQ1: How do air-conditioning and road-traffic noise at different levels affect acoustic appraisal, perceived workload, cognitive performance, and task-period EEG spectral power?

RQ2: Are perceived, cognitive, and EEG responses consistent within participants, that is, do changes in one domain accompany changes in the others?

RQ3: When cognitive performance does not differ significantly between acoustic conditions, can acoustic appraisal, perceived workload, and EEG responses still distinguish between them?

Consistency between domains was treated as an empirical question rather than an assumption. RQ3 concerns conditions that performance measures did not distinguish statistically. Its results are therefore interpreted as showing which responses distinguished these conditions, not as evidence that performance was unaffected or that one type of response is more sensitive than another.

Table 1. Representative sound-exposure studies and the relationships examined across response domains.

| Study and setting | Exposure and activity | Measurements / relationship examined | Boundary for comparison |
|---|---|---|---|
| (Meng et al., 2021) Office experiment | AC, traffic, and other sources; varied levels; multitask visual cognition. | Source/level contrasts in response time and memory accuracy. | Direct AC/RT performance precedent; full subjective-measure inventory not verified. |
| (Pellegatti et al., 2026) Middle-school classrooms | Ventilation, traffic, playground, birdsong, and quiet; reading and calculation. | Soundscape and task outcomes reported in parallel; no EEG outcome in the evidence account. | School-age participants; a common active level rather than the present graded series. |
| (Tsujimura & Ueno, 2010) Anechoic room | No noise, AC, and talking noise during classroom-related tasks. | Subjective disturbance, task correctness, and frontal-midline theta. | Education-related multidomain precedent; not a physical classroom field experiment. |
| (Eqlimi et al., 2020) Lecture analogue | Environmental background sounds during attended lectures. | EEG features related to later examination scores in participant-level models. | Delayed knowledge retention, not same-task PI or condition-wise sound appraisal. |
| (N. Zhang et al., 2025) Listening laboratory | Traffic and spring-water recordings at 40–60 dBA, plus control. | Sound-comfort and pleasure ratings, EEG, ECG; direct rating–power correlations. | No concurrent scored task; repeated-participant adjustment not established in the account. |
| (Mohammadi et al., 2023) Attention experiment | Tone frequency varied with loudness in phon during an attention test. | NASA-TLX, attention metrics, and alpha/gamma ratio; direct correlations. | Frequency–loudness confounding; ordinary correlations do not establish a within-person effect. |
| (Chong et al., 2025) Office experiment | Dialog, music, and office sound at three levels; six tasks. | Subjective, performance, and EEG outcomes; subjective–behavior and level–EEG analyses. | Direct EEG–questionnaire/task pairing not verified; gamma-processing inconsistency. |
| (Rosenkranz et al., 2026) Simulated surgery | Irrelevant speech and operating-room sound under two task demands. | Subjective workload, performance, and neural tracking; participant-level workload–EEG analysis. | Occupational activity and stimulus tracking, not broad task-period spectral power. |
| Present study University classroom | AC/RT recordings at 55/60/65 dBA and control; four tasks. | Appraisal, workload, task outcomes, and EEG; within-participant rating–EEG and same-task PI–EEG pairs. | Exploratory estimates; analysis samples and temporal aggregation differ; no mechanism test. |

# 2. Methods

## 2.1. Overview

This study adopted a within-subject repeated-measures design to examine human responses to environmental noise in classrooms. Each participant completed the same experimental protocol under different type and level noise conditions, which enabled the effects of noise type and sound level to be evaluated within the same participants while reducing between-subject variability.

Responses to the acoustic conditions were assessed from three complementary perspectives: perceived responses, cognitive performance, and neural responses. Perceived responses were obtained using questionnaires that include acoustic perception and perceived workload, providing information on how participants experienced the acoustic environment and task demands. Cognitive performance was derived from a set of predefined cognitive tasks. Neural responses were measured using electroencephalography (EEG) during cognitive task to characterize changes in brain activity that may not be directly evident from perceived or cognitive outcomes. These response domains were analyzed jointly to represent distinct but complementary aspects of the response to environmental noise. The multimodal design therefore allowed the effects of environmental noise in classrooms to be examined across perceived, cognitive, and neural levels and further enabled associations among these response domains to be explored.

## 2.2 Experimental measurement

The experiment comprised acoustic-exposure component and complementary response-measurement components: perceived responses, cognitive-task performance, and neural activity. For each component, the experimental measurement instrument, the recorded data, and the analyze variables are described below.

### 2.2.1 Acoustic stimuli

Two representative sources of environmental noise in classrooms were selected as acoustic stimuli: Air-conditioning (AC) and road-traffic (RT) noise. These sources were chosen because they represent two common but acoustically distinct forms of environmental noise encountered in Hong Kong classrooms. Moreover, each noise source was reproduced at three sound levels of 55, 60, and 65 dBA. The three levels formed a controlled 5-dBA intensity gradient, allowing the effect of increasing sound level and its interaction with noise type to be examined systematically. A control condition without experimentally added noise was additionally included to provide a within-participant reference against which responses under noise exposure could be evaluated.

### 2.2.2 Perceived response

Perceived responses were assessed using two questionnaire instruments targeting participants' experience of the acoustic environment and the workload associated with task completion. Acoustic perception was evaluated using the Classroom Soundscape Assessment (CSA) (Visentin et al., 2023). Three attributes were retained for the present study: loudness, pleasantness, and arousal (rated from calm to exciting), which characterize participants' perceived acoustic experience under different noise conditions. Perceived workload was assessed using the NASA Task Load Index (NASA-TLX) (Hart, 2006) . Participants rated six dimensions of workload: Mental Demand, Physical Demand, Temporal Demand, Performance, Effort, and Frustration. The NASA-TLX was selected to quantify participants perceived workload during different noise conditions. The original questionnaire formats are provided in Appendix A.

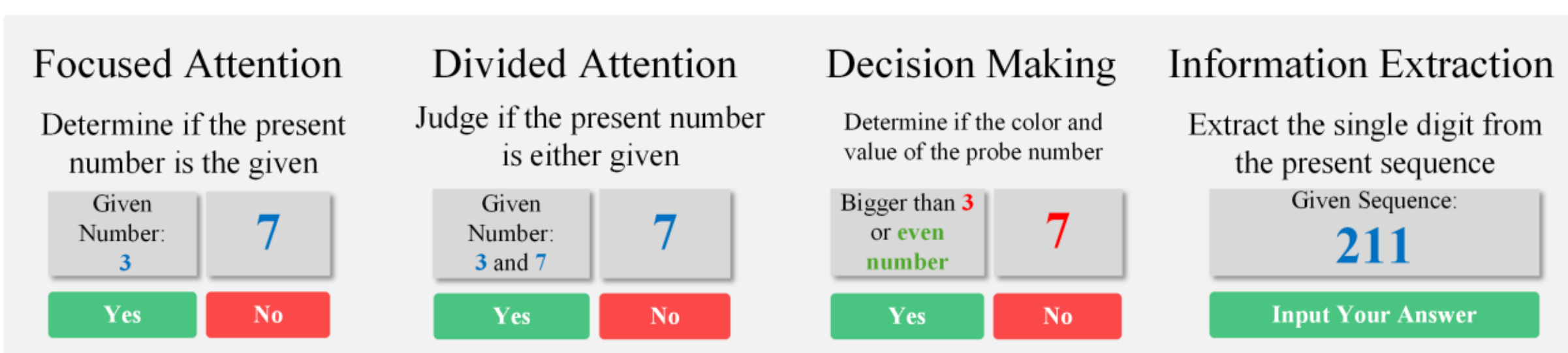

Fig. 1. Cognitive task paradigms used to assess focused attention, divided attention, decision making, and information extraction.

2.2.3 Cognitive performance

Cognitive performance was assessed using four computer-based tasks designed to represent different learning-related cognitive demands, as illustrated in Fig. 1. These cognitive tasks were developed based on established paradigms commonly used to assess attention and executive processing, with task difficulty progressively varied from single-target detection to multi-rule decision making and information extraction. The use of multiple task demands was motivated by previous noise–cognition studies showing that noise effects can depend on the cognitive process being engaged (Bush et al., 2003; Green et al., 2012; Rogers & Monsell, 1995; Sternberg, 1966) . Specifically, Task 1, focused attention, required participants to determine whether a presented numerical stimulus matched a single target. Task 2, divided attention, increased monitoring demands by requiring participants to compare each stimulus with two target numbers. Task 3, decision making, asked participants to integrate multiple stimulus characteristics, including colour, numerical magnitude, and parity, before making a response. Finally, Task 4, information extraction, required participants to identify specified numerical information from a short numerical sequence.

Together, the tasks provided measures spanning focused attention, divided monitoring, cognitive control, and information processing. The task structures remained consistent across acoustic conditions, while individual test items were randomly selected from predefined task-specific question pools to reduce direct repetition of identical stimuli. All tasks were presented using OpenSesame (Mathôt et al., 2012) , an open-source platform for behavioral experiment design and stimulus presentation. The experimental log recorded the participant’s response, response time, task identity, and event timing. These trial-level records were subsequently used to derive task-specific accuracy, response time, and a performance index (PI = Accuracy/response time).

2.2.4 Neural responses

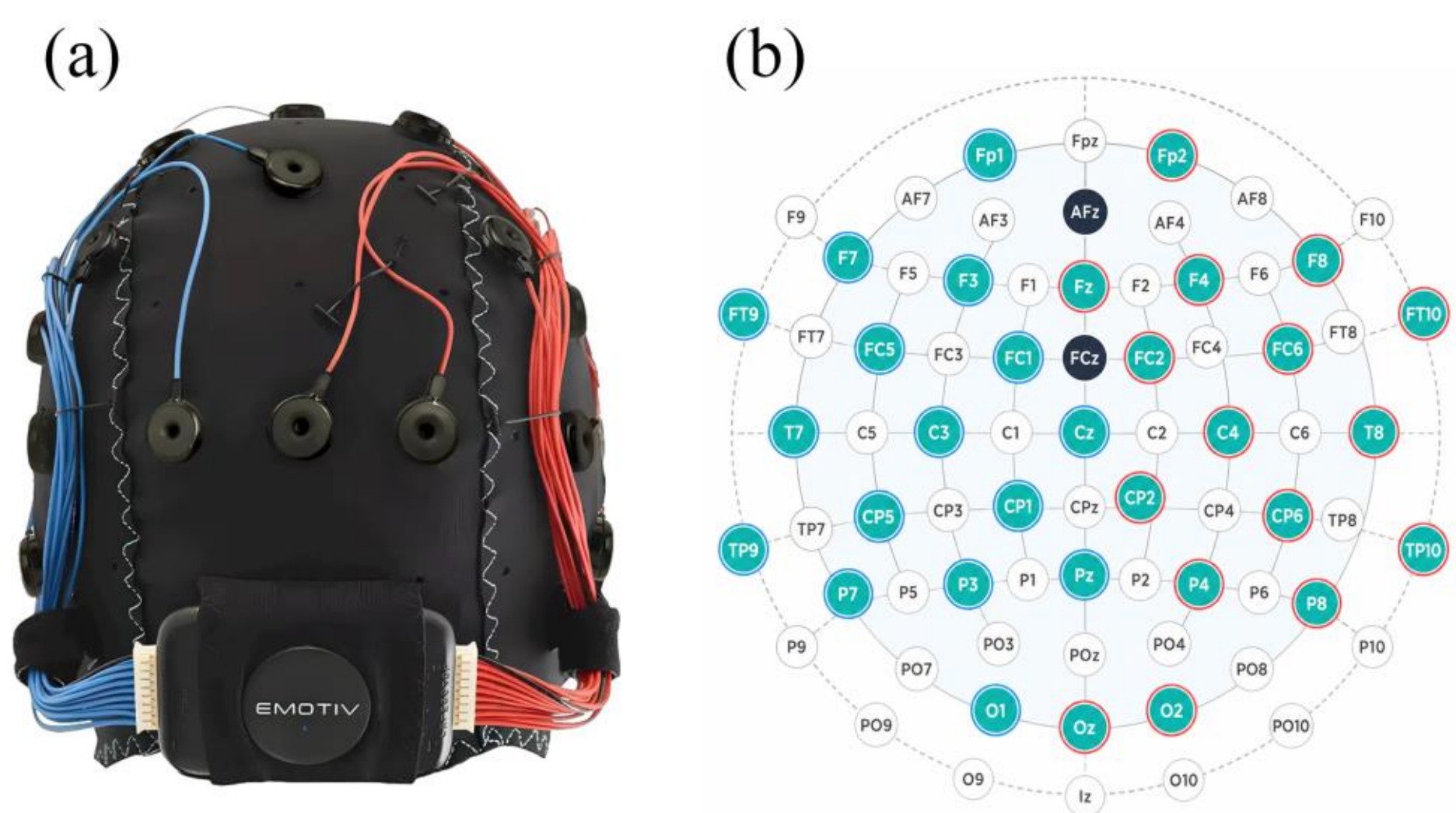

Fig. 2. EEG acquisition system used in the study: (a) Emotiv Flex 2 32-channel wireless EEG headset and (b) corresponding electrode montage(*Flex 2 - 32 Channel Wireless EEG Head Cap System – Emotiv*, n.d.) .

Neural responses were recorded continuously during cognitive-task performance using an Emotiv Flex 2 Saline 32-channel wireless EEG system (Fig. 2)(*Flex 2 - 32 Channel Wireless EEG Head Cap System – Emotiv*, n.d.) . The recording electrodes were arranged using a manufacturer-supported montage based on the international 10–20 system, with two additional control electrodes positioned at TP9 and TP10. EEG signals were sampled at 256 Hz, and the headset remained in place throughout the experimental blocks.

EEG was included to provide a neurophysiological measure complementary to perceived responses and cognitive performance. Changes in neural processing may occur even when participants report similar subjective experiences or maintain comparable task performance across acoustic conditions.

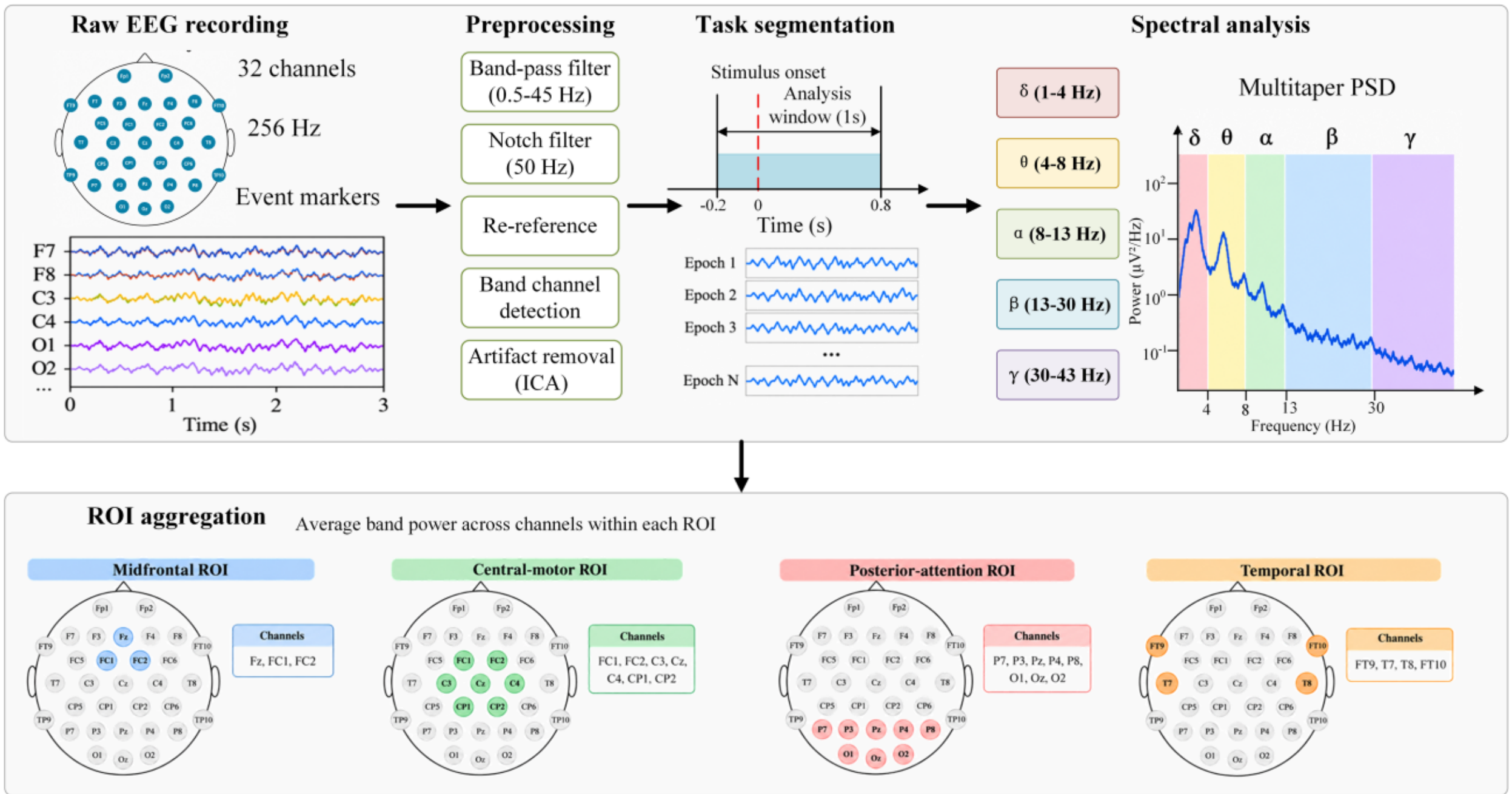

Fig. 3. Workflow for EEG data processing and ROI-based spectral analysis.

EEG acquisition was synchronized with cognitive-task presentation using LabRecorder (*LabRecorder - Secure LSL Documentation*, n.d.), a recording application that enables time-synchronized collection of multiple data streams. Event markers generated by OpenSesame were stored together with the continuous EEG stream, allowing the recordings to be subsequently segmented according to cognitive task and acoustic condition. As shown in Fig. 3, the recorded EEG signals were subsequently preprocessed to reduce recording noise and physiological artifacts, after which the continuous data were segmented into task-specific periods using the synchronized event markers. Valid task segments were divided into short epochs and transformed into spectral-power estimates across five frequency bands: δ, θ, α, β, and γ. To reduce channel-level dimensionality and facilitate functionally interpretable comparisons, spectral power was further aggregated within four predefined functional scalp regions of interest (ROIs): Midfrontal, Central-motor, Posterior-attention, and Temporal, as shown in Table 2. These ROIs were selected to represent neural activity related to cognitive control, reaction, attention, and noise processing, respectively.

Table 2. Functional scalp regions of interest used for EEG spectral-power analysis.

| Functional ROI | Included channels | Primary functional interpretation |
|---|---|---|
| Midfrontal | Fz, FC1, FC2 | Cognitive control and conflict processing (Eisma et al., 2021; McKewen et al., 2021; Nigbur et al., 2011) |
| Central-motor | FC1, FC2, C3, Cz, C4, CP1, CP2 | Response preparation and sensorimotor processing (Boulay et al., 2011) |
| Posterior-attention | P7, P3, Pz, P4, P8, O1, Oz, O2 | Visual processing and attentional allocation (Sokoliuk et al., 2019) |
| Temporal | FT9, T7, T8, FT10 | Auditory/noise-related temporal and frontotemporal activity(Jafari et al., 2019; Manohare et al., 2023) |

#### 2.2.5 Experimental procedure

As presented in Fig. 4, participants were briefed on the experimental procedure and familiarized with the cognitive tasks before data collection. The EEG headset was then fitted, and signal quality was checked before the formal experiment began. Each participant subsequently completed all acoustic conditions in a randomized order. In experiment sessions, EEG activity and cognitive task responses were recorded concurrently throughout task performance. Immediately after task completion, participants completed the questionnaires assessing their perception of the acoustic environment and perceived workload, followed by a rest interval before proceeding to the next acoustic condition. This block-wise procedure ensured that perceived responses, cognitive performance, and EEG activity could be matched to the same participant and acoustic condition.

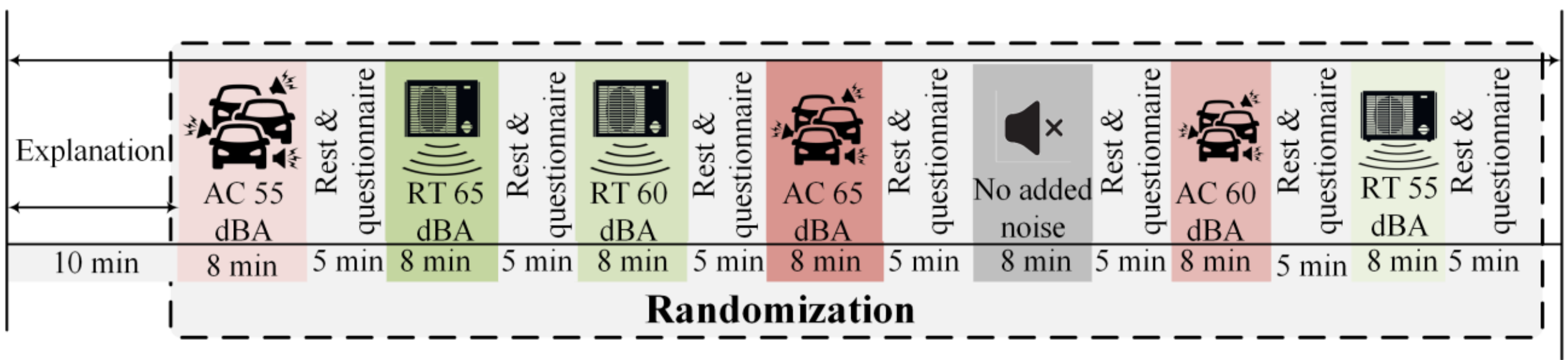

Fig. 4. Experimental procedure and sequential collection of perceived, cognitive, and neural response data.

## 2.3 Participants and experimental setting

### 2.3.1 Participant selection

University students aged 18–30 years were recruited in Hong Kong. Participants were eligible if they reported normal hearing and no known neurological disorders or cognitive impairments. A total of 22 participants completed the experimental protocol. Following EEG data-quality screening, 16 participants were retained for the final analysis, including 11 males and 5 females. To ensure that perceived responses, cognitive performance, and neural responses were compared using the same participant sample, subsequent multimodal analyses were based on these 16 participants. Participant data were anonymized throughout data processing and analysis. Selected participants performing the cognitive tasks while wearing the EEG headset are shown in Fig. 5, with identifiable facial information obscured for privacy.

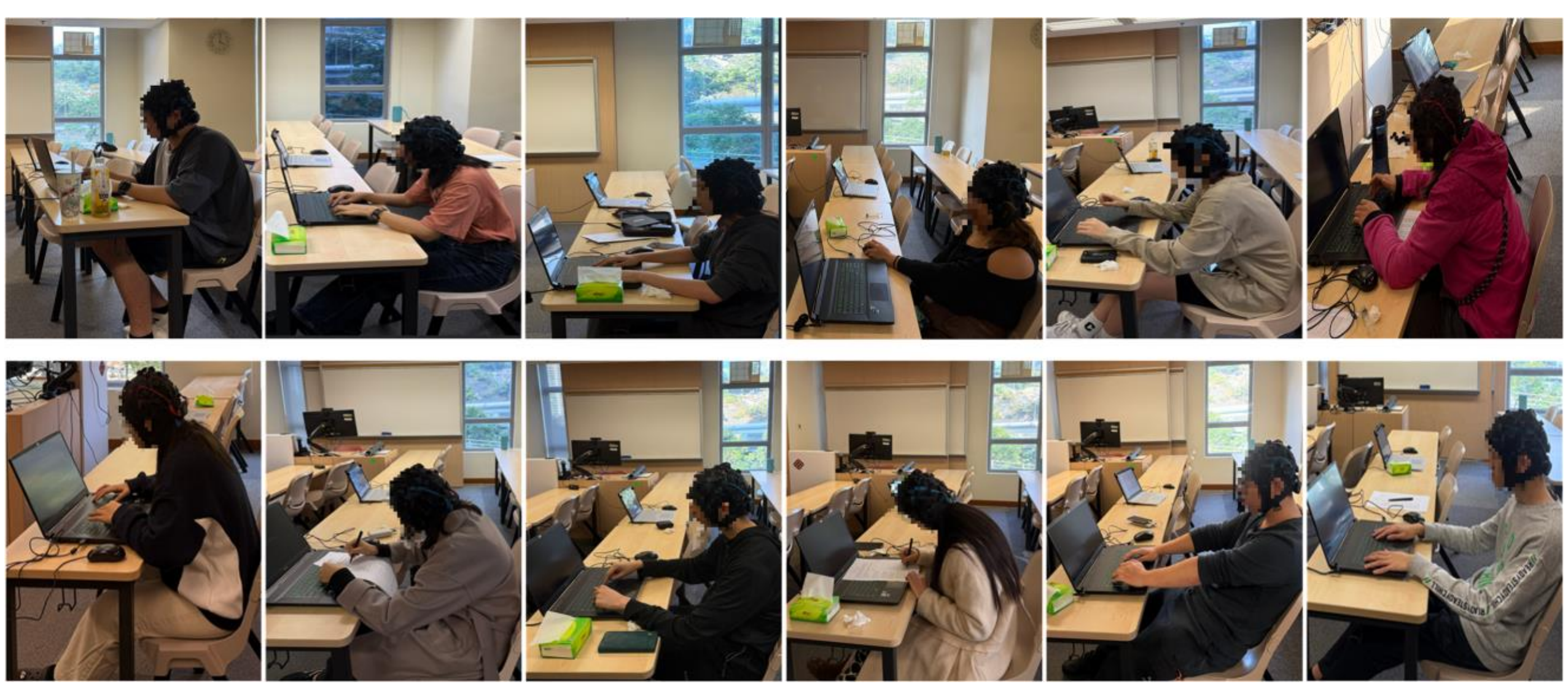

Fig. 5. Selected participants completing the cognitive tasks while wearing the EEG headset during the experiment. Identifiable facial information was anonymized to protect participant privacy.

2.3.2 Experiment setting

The Experimental environment and measurement configuration are demonstrated in Fig. 6. Specifically, the experiment was conducted in a teaching classroom in the Z Core of The Hong Kong Polytechnic University, Hong Kong. The classroom retained the spatial and visual characteristics of a conventional university teaching environment. During data collection, the windows remained closed to reduce uncontrolled outdoor sound and to maintain consistent acoustic conditions across experimental blocks. AC and RT noise were reproduced at three sound levels—55, 60, and 65 dBA—resulting in six active-noise conditions. A control condition with no experimentally added noise was also included, under which participants completed the same experimental protocol in the existing classroom background environment.

Acoustic stimuli were reproduced using a Sony SRS-XE200/LC wireless speaker positioned near the classroom window and kept at the same location throughout the experiment. The position was selected to approximate the direction from which environmental noise would typically be perceived in the classroom. Before each active-noise block, the loudspeaker output was adjusted to the corresponding target sound level, as shown in Fig. 6(c). The existing window-mounted air-conditioning unit in the classroom was not used to generate the experimental AC stimulus. In the control condition, no experimental recording was played, while the same cognitive-task and measurement procedures were retained.

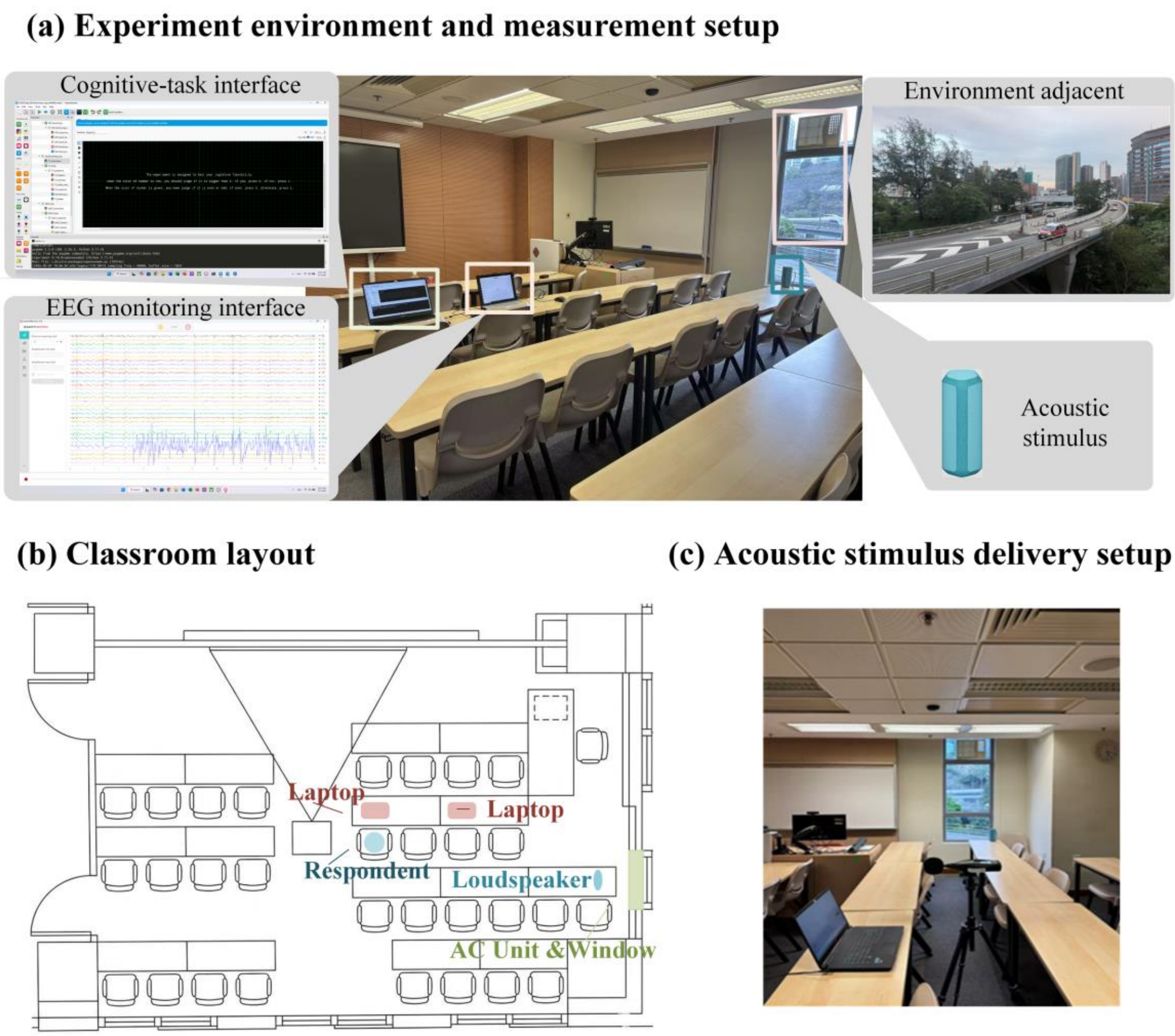

Fig. 6. Experimental environment and measurement configuration: (a) classroom environment and measurement interfaces; (b) spatial layout of the experiment; and (c) in-situ acoustic stimulus delivery setup.

### 2.4. Data analysis

The data analysis comprised three stages: (1) processing the three response data and deriving the corresponding analytical indicators, (2) evaluating the effects of acoustic conditions within each response domain, and (3) examining cross-domain associations between perceived, cognitive, and neural responses. Exact $p$ values, effect sizes, and 95% confidence intervals were reported so that each result could be weighed against the number of tests performed.

2.4.1 Stage 1: Data processing

**Perceived responses** For perceived responses, the three CSA ratings—loudness, pleasantness, and arousal—were retained as separate acoustic-perception outcomes. For NASA-TLX, the six dimensions (Mental Demand, Physical Demand, Temporal Demand, Performance, Effort, and Frustration) were converted to numerical scores. The Raw TLX was calculated as the arithmetic mean of the six dimensions, while the Weighted TLX incorporated the dimension-specific pairwise-comparison weights. Raw and Weighted TLX were used as the overall workload outcomes, with the six individual dimensions retained as supporting measures.

**Cognitive performance** Cognitive-task logs were organized by acoustic condition, participant, and task. Trial-level correctness and response time were aggregated within each Participant × Condition × Task combination to obtain accuracy and mean response time. A performance index (PI) was further calculated to integrate response accuracy and speed:

$$PI = \frac{Accuracy}{Response\ Time}. \tag{1}$$

PI was used as the primary cognitive performance indicator, while accuracy and response time were retained as complementary outcomes for interpretation. The four cognitive tasks were processed and analyzed separately because they represented different cognitive demands.

**Neural responses** EEG processing followed the workflow summarized in Fig. 3. After preprocessing and artifact removal, the continuous recordings were segmented into the four cognitive-task periods using the synchronized event markers. Valid task segments were divided into non-overlapping 1-s epochs, and epochs failing the predefined signal-quality criteria were excluded. Power spectral density was estimated using the multitaper method (Thomson, 1982) , and absolute power was extracted for five frequency bands: δ (1–4 Hz), θ (4–8 Hz), α (8–13 Hz), β (13–30 Hz), and γ (30–45 Hz). Band power was first averaged across the electrodes within each predefined ROI and then across valid epochs. To express noise-related changes relative to each participant's own reference condition, the ROI-level band power under each active-noise condition was normalized to the matched control value:

$$\Delta P(\%) = \frac{P_{\text{noise}} - P_{\text{control}}}{P_{\text{control}}} \times 100\%. \tag{2}$$

Positive values indicated greater spectral power under noise exposure relative to the control condition, whereas negative values indicated lower power. Hence, the derived variables used in the subsequent analyses are summarized in Table 3.

Table 3. Summary of derived outcome variables across the three response domains.

| Measurement domain | Measurement | Recorded variables |
|---|---|---|
| Perceived response | CSA | Loudness, pleasantness, arousal |
| | NASA-TLX | Raw TLX, Weighted TLX, Mental Demand, Physical Demand, Temporal Demand, Performance, Effort, and Frustration |
| Cognitive task performance | OpenSesame | Accuracy, response time, performance index |
| Neural response | Emotiv Flex 2 Saline | $\delta$, $\theta$, $\alpha$, $\beta$, and $\gamma$ bands within the Midfrontal, Central-motor, Posterior-attention, and Temporal ROIs |

#### 2.4.2 Stage 2: Statistical analysis

**Perceived and cognitive response** For descriptive presentation, cognitive outcomes were summarized using mean values and visualized with boxplots to illustrate their distributions across acoustic conditions. Furthermore, perceived and cognitive outcomes were analyzed using a two-stage within-subject strategy. The overall effect of experimentally added noise was evaluated by comparing each participant's control-condition value with that participant's mean value across the active-noise conditions. The active-noise conditions were then analyzed using a repeated-measures factorial design, with Noise Type (AC vs. RT) and Noise Level (55, 60, and 65 dBA) as within-subject factors. Separate analyses were conducted for loudness, pleasantness, arousal, Raw TLX, Weighted TLX, and the six NASA-TLX dimensions. For cognitive performance, analyses were conducted separately for each of the four tasks, with PI treated as the primary outcome and accuracy and response time used as complementary indicators. For noise level and the type × level interaction, which have two numerator degrees of freedom, $p$ values were Greenhouse–Geisser corrected. Statistical significance was defined as $p < .05$.

**Neural response** Following (Mir et al., 2022) **,** linear mixed-effects models were used to analyze the EEG responses because this method can accommodate repeated observations across noise types, noise levels, and frequency bands while accounting for participant-level variability. Separate models were fitted for each Cognitive Task-ROI combination, with the control-normalized percentage change in spectral power as the dependent variable. Noise Type, Noise Level, Frequency Band, and their two- and three-way interactions were entered as categorical fixed effects, while Participant was included as a random intercept:

$$\Delta P \sim Noise\,Type \times Noise\,Level \times Frequency\,Band + (1 \mid Participant). \qquad (3)$$

Overall fixed effects were evaluated using Wald $\chi^2$ tests. Significant interactions were hierarchically decomposed using estimated marginal means and simple-effect comparisons. Effect estimates were reported with 95% confidence intervals.

#### 2.4.3 Stage 3: Cross-domain association

Cross-domain associations were examined using repeated-measures correlation, which estimates within-participant associations while accounting for repeated observations across acoustic conditions. Two sets of associations were examined. First, perceived–neural associations were evaluated between the selected perceived-response measures (Weighted TLX, loudness, pleasantness, and arousal) and relative EEG band power within each ROI. Second, cognitive–neural associations were evaluated between task-specific PI and the corresponding relative EEG band power within each ROI.

The correlations were computed across the six noise conditions using relative band power, defined as the power in each band divided by the total power between 1 and 45 Hz. The questionnaire ratings were paired with the EEG recorded in the same condition, and the PI of each task was paired with the EEG recorded during that task. Correlations were computed separately for each ROI. For each test, the repeated-measures correlation coefficient ($r_{\mathrm{rm}}$), its 95% confidence interval, and the $p$ value were reported.

## 3. Results

This section first presents the effects of the acoustic conditions on perceived responses, cognitive performance, and EEG spectral power, followed by the within-participant associations between EEG and the other responses.

### 3.1 Perceived responses

#### 3.1.1 Acoustic appraisal

Fig. 7 presents the loudness, pleasantness, and arousal ratings across the control and noise conditions. Compared with the control condition, all six noise conditions were rated as louder, less pleasant, and more arousing. Loudness ratings increased with sound level for both noise types, with a steeper increase for RT noise; RT noise was rated slightly quieter than AC noise at 55 dBA but louder at 60 and 65 dBA. Pleasantness ratings decreased as the sound level increased, whereas arousal ratings increased with sound level under RT noise and, to a lesser extent, under AC noise.

The results of the two-stage analysis are summarized in Table 4. Compared with the control condition, the mean of the six noise conditions was rated significantly louder ($\Delta$=5.77, $t$(15)=6.51, $p$<.001), less pleasant ($\Delta$=−4.03, $t$(15)=−3.26, $p$ =.005), and more arousing ($\Delta$=4.11, $t$(15)=4.89, $p$<.001). Among the noise conditions, noise level significantly affected all three ratings, namely loudness ($F$(2, 30)=11.72, $p$<.001, $\eta_p^2$=.439), pleasantness ($F$(2, 30)=7.86, $p$ =.004, $\eta_p^2$ =.344), and arousal ($F$(2, 30)=5.41, $p$ =.010, $\eta_p^2$=.265). Noise type affected only loudness, with RT noise rated louder than AC noise on average ($F$(1, 15)=6.62, $p$=.021, $\eta_p^2$=.306), and loudness also showed a significant type × level interaction ($F$(2, 30)=6.89, $p$=.004, $\eta_p^2$=.315), consistent with its steeper increase under RT noise. No other type or interaction effect was significant (all $p$≥.05). Acoustic appraisal therefore responded to both the presence and the level of noise, whereas the two noise types differed only in loudness.

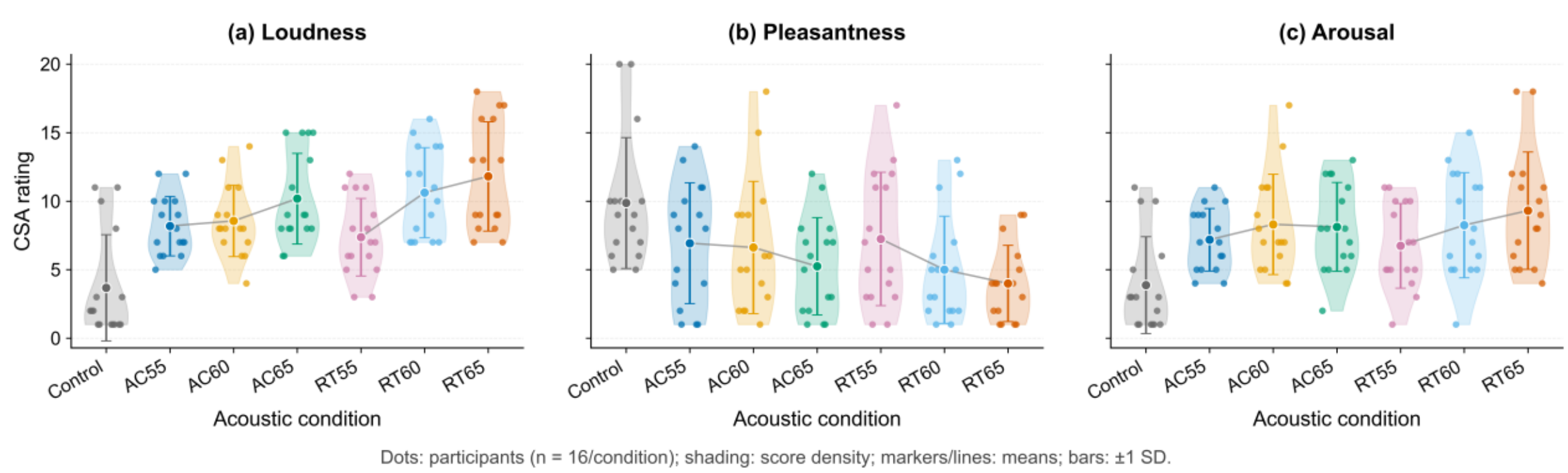

Fig. 7. Acoustic appraisal ratings of (a) loudness, (b) pleasantness, and (c) arousal (calm–exciting) across the control and six noise conditions. Points represent means, and error bars represent ±1 $SD$ ($N$ = 16).

Table 4. Results of the two-stage within-participant analysis of acoustic appraisal ratings.

| Rating | Noise vs control | Noise type | Noise level | Type × level |
|---|---|---|---|---|
| Loudness | Δ = 5.77, $t(15) = 6.51$<br>$p < .001$ | $F(1, 15) = 6.62$<br>$p = .021$, $\eta_p^2 = .306$ | $F(2, 30) = 11.72$<br>$p < .001$, $\eta_p^2 = .439$ | $F(2, 30) = 6.89$<br>$p = .004$, $\eta_p^2 = .315$ |
| Pleasantness | Δ = −4.03, $t(15) = -3.26$<br>$p = .005$ | $F(1, 15) = 1.51$<br>$p = .239$, $\eta_p^2 = .091$ | $F(2, 30) = 7.86$<br>$p = .004$, $\eta_p^2 = .344$ | $F(2, 30) = 3.39$<br>$p = .052$, $\eta_p^2 = .184$ |
| Arousal | Δ = 4.11, $t(15) = 4.89$<br>$p < .001$ | $F(1, 15) = 0.15$<br>$p = .702$, $\eta_p^2 = .010$ | $F(2, 30) = 5.41$<br>$p = .010$, $\eta_p^2 = .265$ | $F(2, 30) = 1.94$<br>$p = .165$, $\eta_p^2 = .115$ |

Note: Δ = mean of the six noise conditions minus the control condition (0–20 scale).

### 3.1.2 Perceived workload

Figs. 8 and 9 present the overall NASA-TLX scores and the six workload dimensions, respectively. Both raw and weighted TLX scores were higher under all six noise conditions than under the control condition, with the highest mean scores observed under RT noise at 60 dBA. A similar tendency was found for most individual dimensions, although the variability between participants was considerable.

As shown in Table 5, raw and weighted TLX scores were significantly higher under noise than under the control condition, by 9.59 points ($t(15) = 2.47$, $p = .026$) and 10.17 points ($t(15) = 2.54$, $p = .022$), respectively. Of the individual dimensions, temporal demand (Δ = 13.54, $t(15) = 2.41$, $p = .029$) and effort (Δ = 8.59, $t(15) = 2.16$, $p = .047$) were also significantly higher under noise, whereas mental demand, physical demand, performance, and frustration did not differ significantly ($p = .053$–$.213$). Among the noise conditions, neither raw nor weighted TLX varied significantly with noise type, noise level, or their interaction (all $p \geq .316$). At the dimension level, mental demand was higher under RT than under AC noise ($F(1, 15) = 5.30$, $p = .036$, $\eta_p^2 = .261$), and physical demand showed a significant type × level interaction ($F(2, 30) = 5.09$, $p = .018$, $\eta_p^2 = .253$). Perceived workload was therefore higher under noise than under the control condition but, apart from these two dimensions, did not depend on the type or level of noise.

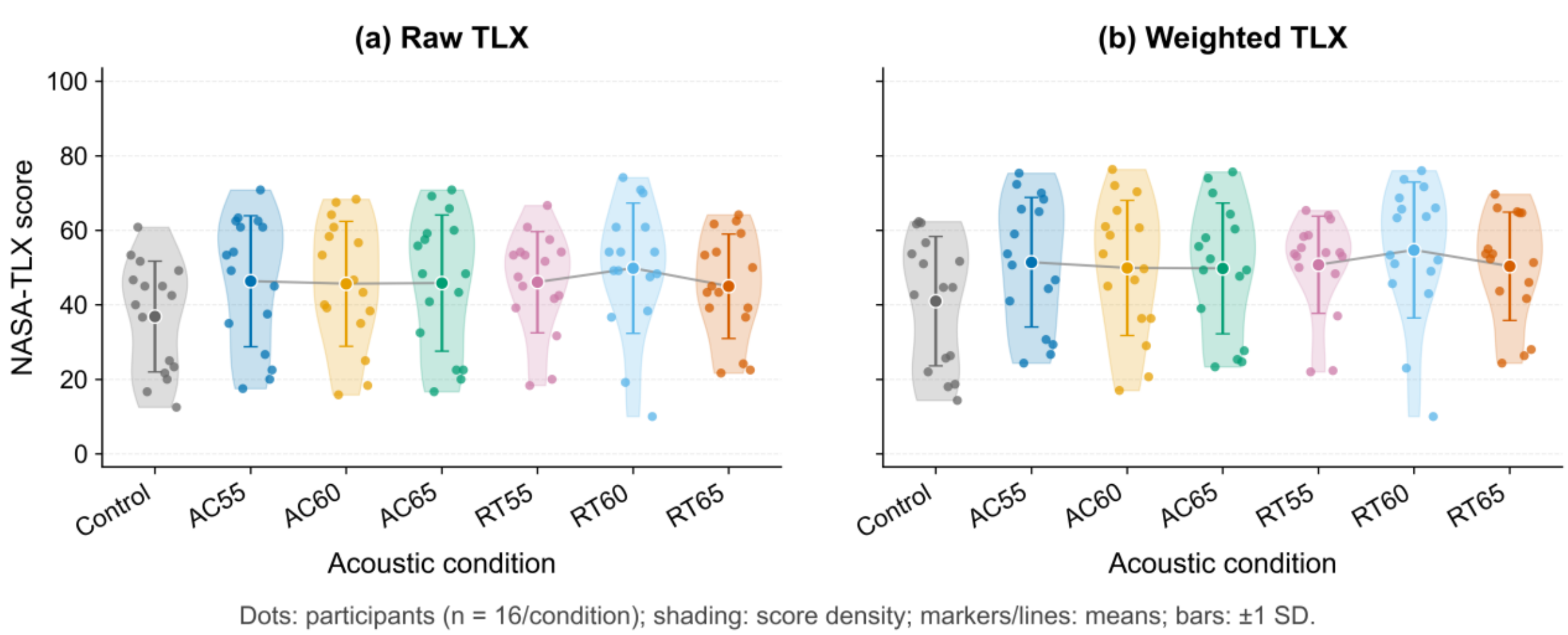

Fig. 8. (a) Raw and (b) weighted NASA-TLX scores across the control and six noise conditions. Points represent means, and error bars represent ±1 *SD*.

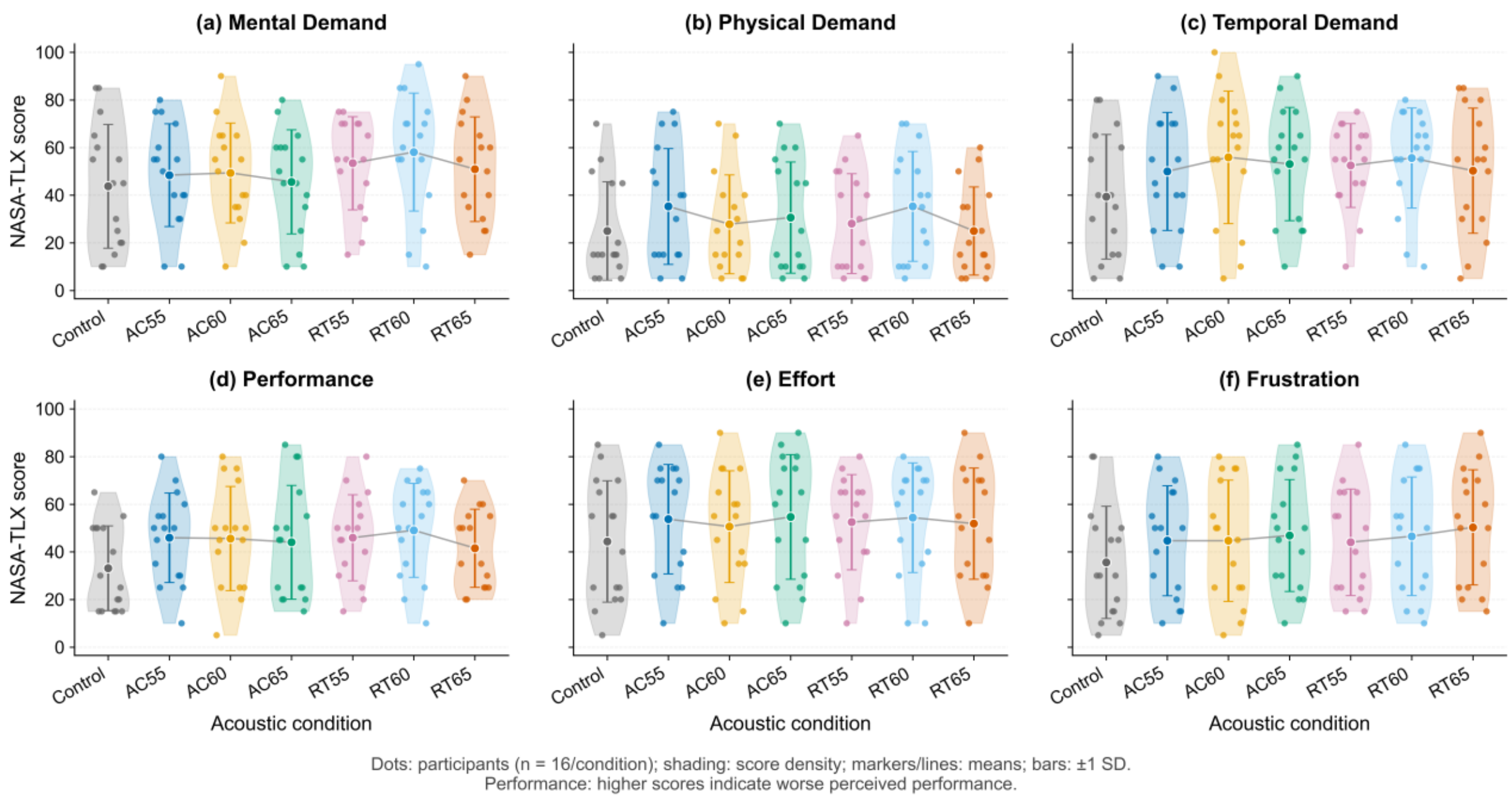

Fig. 9. Ratings of the six NASA-TLX workload dimensions across the control and six noise conditions. Points represent means, and error bars represent ±1 *SD*.

Table 5. Results of the two-stage within-participant analysis of NASA-TLX scores.

| Score | Noise vs control | Noise type | Noise level | Type × level |
|---|---|---|---|---|
| Raw TLX | $\Delta = 9.59$, $t(15) = 2.47$<br>$p = .026$ | $F(1, 15) = 0.42$<br>$p = .528$, $\eta_p^2 = .027$ | $F(2, 30) = 1.17$<br>$p = .316$, $\eta_p^2 = .072$ | $F(2, 30) = 1.01$<br>$p = .357$, $\eta_p^2 = .063$ |
| Weighted TLX | $\Delta = 10.17$, $t(15) = 2.54$<br>$p = .022$ | $F(1, 15) = 0.71$<br>$p = .414$, $\eta_p^2 = .045$ | $F(2, 30) = 0.90$<br>$p = .396$, $\eta_p^2 = .057$ | $F(2, 30) = 0.88$<br>$p = .395$, $\eta_p^2 = .055$ |
| Mental demand | $\Delta = 7.24$, $t(15) = 1.50$<br>$p = .155$ | $F(1, 15) = 5.30$<br>$p = .036$, $\eta_p^2 = .261$ | $F(2, 30) = 3.03$<br>$p = .069$, $\eta_p^2 = .168$ | $F(2, 30) = 0.19$<br>$p = .804$, $\eta_p^2 = .013$ |
| Physical demand | $\Delta = 5.36$, $t(15) = 1.45$<br>$p = .168$ | $F(1, 15) = 1.15$<br>$p = .300$, $\eta_p^2 = .071$ | $F(2, 30) = 2.06$<br>$p = .147$, $\eta_p^2 = .121$ | $F(2, 30) = 5.09$<br>$p = .018$, $\eta_p^2 = .253$ |
| Temporal demand | $\Delta = 13.54$, $t(15) = 2.41$<br>$p = .029$ | $F(1, 15) = 0.00$<br>$p = .947$, $\eta_p^2 = .000$ | $F(2, 30) = 1.40$<br>$p = .262$, $\eta_p^2 = .086$ | $F(2, 30) = 0.20$<br>$p = .721$, $\eta_p^2 = .013$ |
| Performance | $\Delta = 12.24$, $t(15) = 2.10$<br>$p = .053$ | $F(1, 15) = 0.01$<br>$p = .930$, $\eta_p^2 = .001$ | $F(2, 30) = 1.21$<br>$p = .305$, $\eta_p^2 = .075$ | $F(2, 30) = 0.42$<br>$p = .659$, $\eta_p^2 = .027$ |
| Effort | $\Delta = 8.59$, $t(15) = 2.16$<br>$p = .047$ | $F(1, 15) = 0.00$<br>$p = .963$, $\eta_p^2 = .000$ | $F(2, 30) = 0.09$<br>$p = .894$, $\eta_p^2 = .006$ | $F(2, 30) = 1.14$<br>$p = .322$, $\eta_p^2 = .071$ |
| Frustration | $\Delta = 10.57$, $t(15) = 1.30$<br>$p = .213$ | $F(1, 15) = 0.28$<br>$p = .604$, $\eta_p^2 = .018$ | $F(2, 30) = 0.83$<br>$p = .418$, $\eta_p^2 = .052$ | $F(2, 30) = 0.23$<br>$p = .785$, $\eta_p^2 = .015$ |

Note: Δ = mean of the six noise conditions minus the control condition (0–100 scale).

### 3.2 Cognitive performance

Fig. 10 presents the distributions of accuracy, mean response time, and PI for the four cognitive tasks. Accuracy approached the ceiling in Tasks 1 and 2, whereas it was lower and more variable in Task 4, and the longest response times were observed in Task 3. Within each task, however, the distributions largely overlapped across the seven acoustic conditions.

Consistent with these descriptive patterns, PI did not differ significantly between the noise conditions and the control condition in any task (Tasks 1–4: $t(15) = 0.06$, 1.48, 1.43, and 0.73; $p = .954$, .159, .173, and .478, respectively). Among the noise conditions, none of the tests of noise type, noise level, or their interaction on PI was significant, the smallest *p* value being that for noise type in Task 2 ($p = .063$). Likewise, accuracy and mean response time did not differ significantly in any of the eight comparisons with the control condition or the tests among the noise conditions. Overall, no significant differences in cognitive performance were found across the tested acoustic conditions.

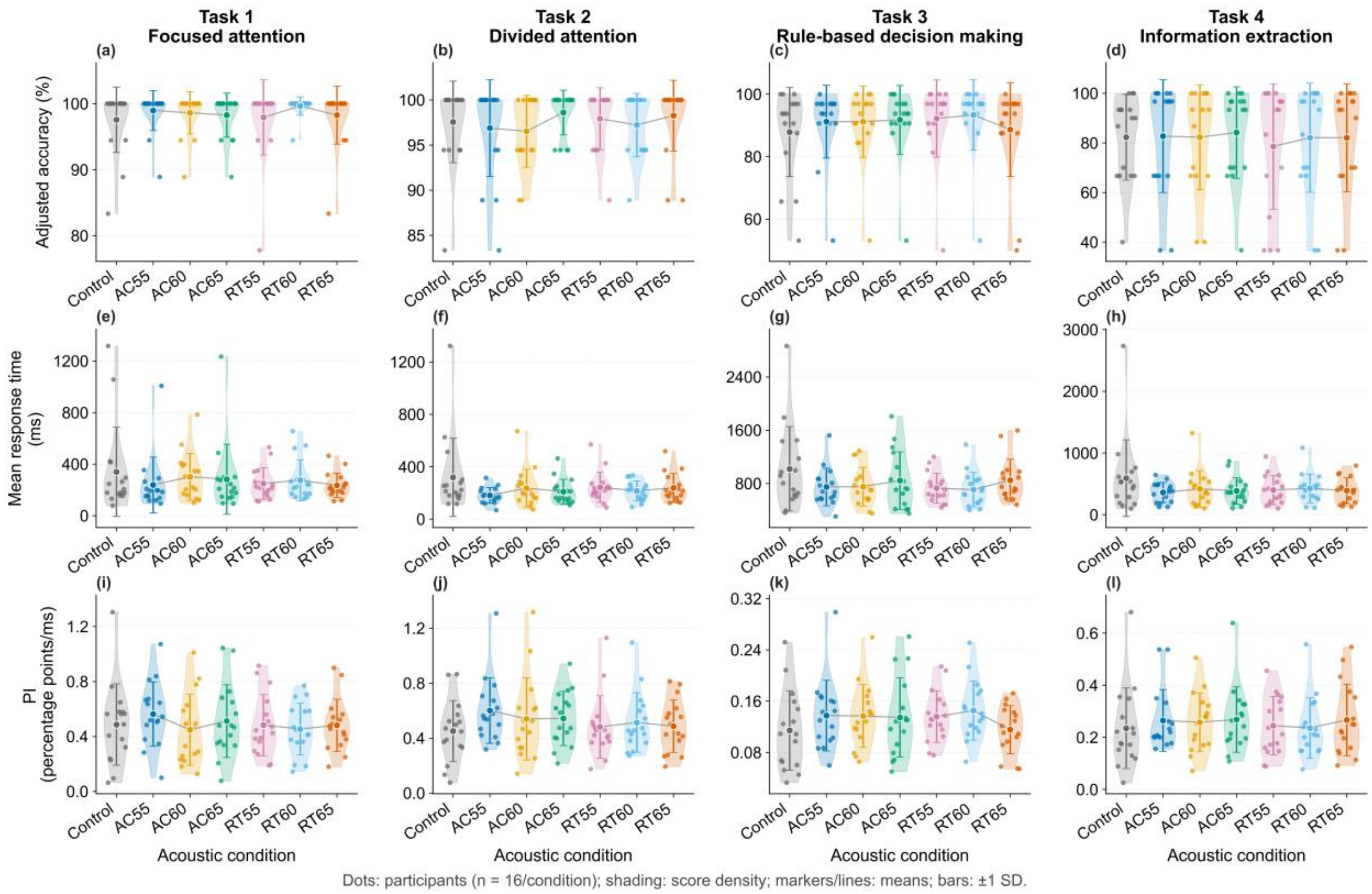

Fig. 10. Distributions of (top) adjusted task accuracy, (middle) mean response time, and (bottom) performance index for the four cognitive tasks across the control and six noise conditions. Accuracy values below 35% were reversed (100 − accuracy) before the PI was calculated, and mean response time includes all formal trials (Section 2.4.1). Boxes indicate medians and interquartile ranges, diamonds indicate means, and grey dots indicate individual participants.

## 3.3 EEG spectral responses

### 3.3.1 Spectral characteristics

Fig. 11 illustrates the task-period power spectral density (PSD) averaged over all 32 channels under the six noise conditions. For all four tasks, the PSD exhibited the typical decrease in power with increasing frequency, and the spectra of the six noise conditions largely overlapped within their 95% confidence intervals. Some descriptive differences could nevertheless be observed, such as the relatively higher power under AC noise at 55 dBA in Task 1 and the lower power under RT noise at 60 dBA in Task 2. Because visual differences between spectra do not indicate statistical significance, the effects of noise type and level were further examined using the ROI-based linear mixed-effects models described in Section 2.4.2.

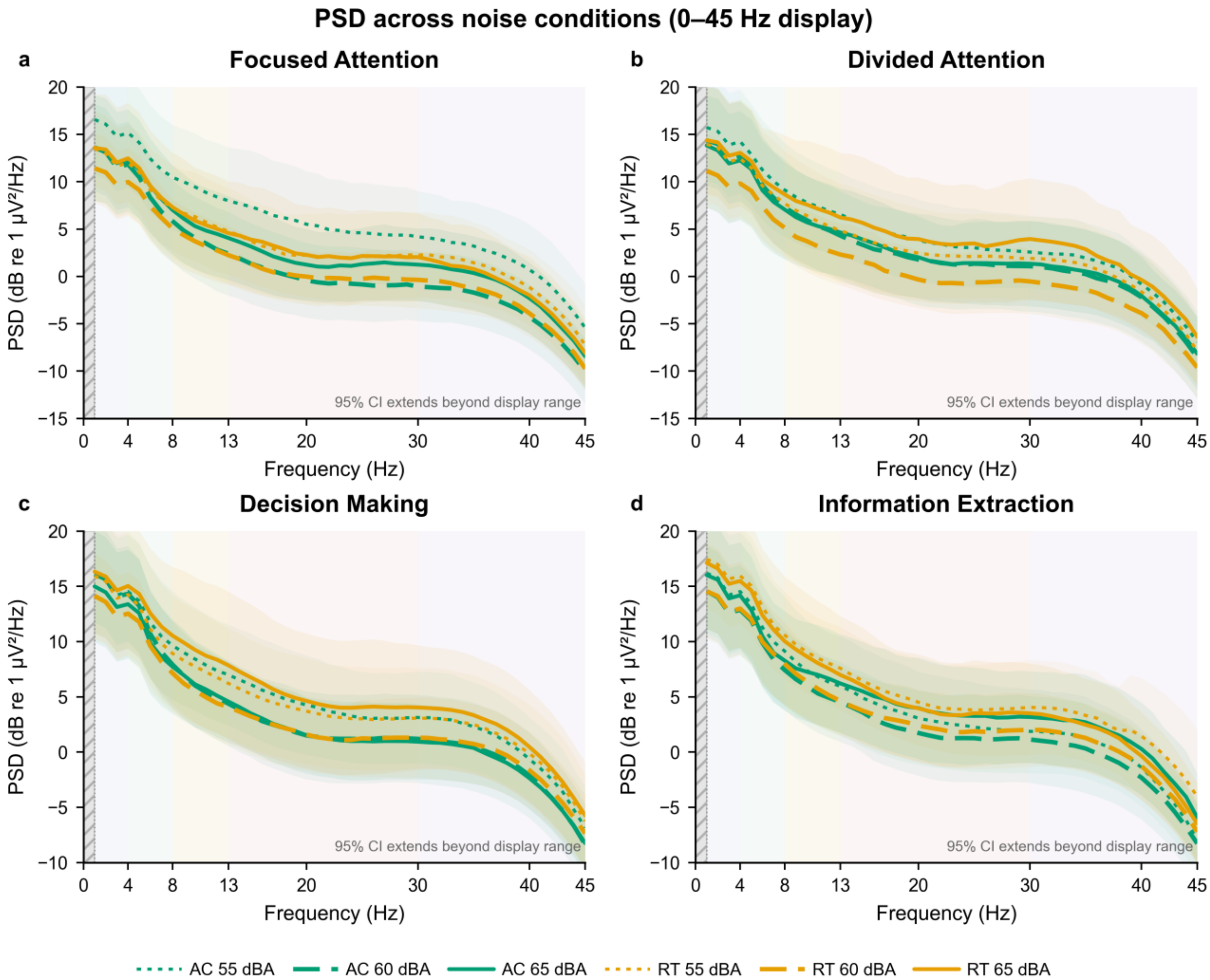

Fig. 11. Task-period PSD averaged over 32 channels under the six noise conditions for the four cognitive tasks. Lines represent the means of participant-level spectra, and shaded areas represent pointwise 95% confidence intervals; background shading marks the frequency bands, and the hatched 0–1 Hz region was excluded from analysis.

### 3.3.2 Effects of noise type and level

Fig. 12 summarizes the omnibus Wald tests of the linear mixed-effects models, in which the normalized change in band power was the dependent variable. Of all the omnibus tests, six main effects of frequency band, four main effects of noise type, six main effects of noise level, and five type × level interactions were significant (detailed in Appendix B). No interaction involving frequency band was significant in any model, indicating that the noise-related changes did not differ significantly between bands. For the significant band main effects, post-hoc comparisons showed larger normalized changes in the delta and theta bands than in the beta and gamma bands; these comparisons concern the normalized change rather than absolute band power. The contrasts that characterize the significant noise-related effects are described below for each task, with differences expressed in percentage points of the normalized change; all contrasts are listed in Appendix C.

**Task 1 (focused attention).** Significant type × level interactions were found in all four ROIs ($\chi^2(2) = 9.27$–$24.49$, all $p \leq .010$), indicating that the difference between AC and RT noise depended on the sound level. In the posterior-attention ROI, the normalized change was larger under AC than under RT noise at 55 dBA (82.38, 95% CI [38.14, 126.61], $p < .001$) and 65 dBA (53.23, 95% CI [9.52, 96.94], $p = .017$) but smaller

at 60 dBA (−67.33, 95% CI [−111.71, −22.95], $p = .003$). A similar reversal occurred in the midfrontal ROI, where the change was larger under RT noise at 55 dBA (−71.67, 95% CI [−118.10, −25.24], $p = .002$) but larger under AC noise at 65 dBA (54.58, 95% CI [8.69, 100.48], $p = .020$). In the central-motor and temporal ROIs, the change was larger under AC noise only at 65 dBA (96.13, 95% CI [49.38, 142.88], $p < .001$) and 55 dBA (73.56, 95% CI [30.12, 117.00], $p < .001$), respectively. The main effects of noise type (central-motor and temporal ROIs) and noise level (central-motor ROI) in this task are therefore qualified by these interactions.

**Task 2 (divided attention).** Noise level significantly affected the central-motor ($\chi^2(2) = 6.46$, $p = .040$) and temporal ($\chi^2(2) = 6.29$, $p = .043$) ROIs. In the central-motor ROI, the change was larger at 65 dBA than at 60 dBA (−45.76, 95% CI [−81.05, −10.47], $p = .011$). In the temporal ROI, the change at 60 dBA was smaller than that at 55 dBA (41.97, 95% CI [5.78, 78.17], $p = .023$) and 65 dBA (−35.16, 95% CI [−69.40, −0.93], $p = .044$).

**Task 3 (decision making).** Noise level significantly affected the posterior-attention ($\chi^2(2) = 17.54$, $p < .001$) and central-motor ($\chi^2(2) = 7.40$, $p = .025$) ROIs, and noise type significantly affected the temporal ROI ($\chi^2(1) = 6.02$, $p = .014$), where the change was larger under RT than under AC noise (−35.73, 95% CI [−64.26, −7.20], $p = .014$). In the posterior-attention ROI, which also showed a significant type × level interaction ($\chi^2(2) = 6.20$, $p = .045$), the level effect was confined to RT noise: the change under RT noise was larger at 65 dBA than at 55 dBA (−101.53, 95% CI [−156.84, −46.22], $p < .001$) and 60 dBA (−108.41, 95% CI [−161.09, −55.72], $p < .001$), whereas no level contrast under AC noise was significant ($p \geq .062$). In the central-motor ROI, the change was larger at 55 dBA than at 60 dBA (45.36, 95% CI [11.03, 79.69], $p = .010$).

**Task 4 (information extraction).** Noise type significantly affected the midfrontal ROI ($\chi^2(1) = 7.93$, $p = .005$), where the change was larger under RT than under AC noise (−42.62, 95% CI [−72.29, −12.95], $p = .005$). Noise level significantly affected the temporal ROI ($\chi^2(2) = 8.32$, $p = .016$), where the change at 65 dBA was larger than that at 55 dBA (−43.33, 95% CI [−77.06, −9.61], $p = .012$) and 60 dBA (−41.06, 95% CI [−73.69, −8.43], $p = .014$).

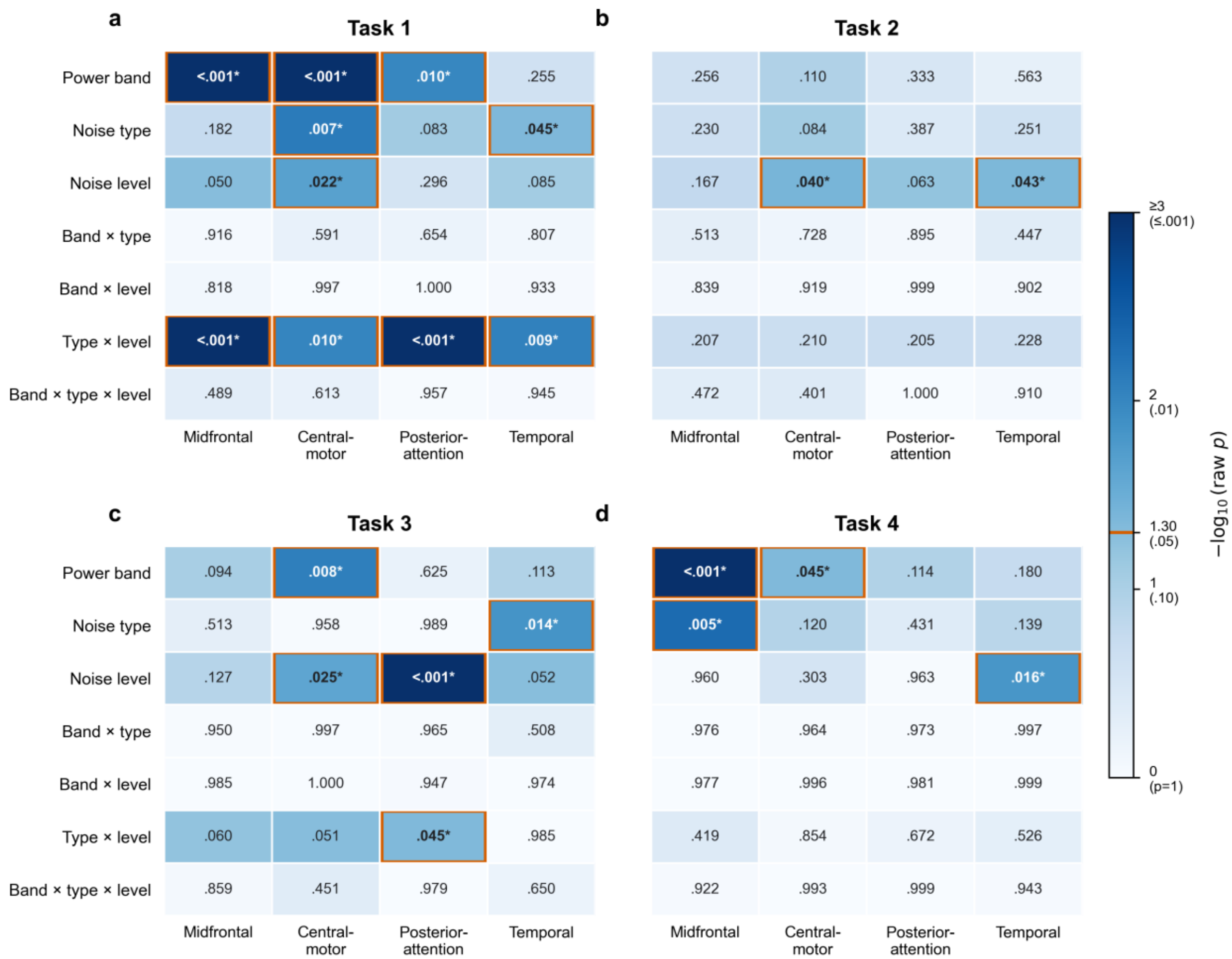

Fig. 12. Omnibus Wald test results of the linear mixed-effects models for the normalized band power in each task and ROI. Cell values indicate uncorrected $p$ values, outlined cells indicate $p < .05$, and colours represent $-\log_{10}(p)$ rather than effect size or direction.

## 3.4 Cross-domain associations

Repeated-measures correlations were used to examine whether relative EEG band power covaried with the questionnaire ratings and with same-task PI within participants across the six noise conditions.

### 3.4.1 Associations between perceived responses and EEG

Fig. 13 presents the correlations between the four questionnaire measures and relative band power in the four ROIs. Of all the correlations, over a quarter is significant, compared with about four expected by chance at $\alpha = .05$. The significant associations involved loudness (11) and pleasantness (9) in all four ROIs and arousal (2) in the temporal ROI, whereas none involved weighted TLX. Specifically, higher loudness ratings were associated with higher relative theta and alpha power and lower relative gamma power, whereas higher pleasantness ratings were associated with lower relative theta power and higher relative gamma power. The strongest associations were found between loudness and gamma power in the central-motor ROI ($r_{rm}$=−.547, 95% CI [−.70, −.35], $p < .001$) and in the posterior-attention and temporal ROIs (both $r_{rm} = -.535$, $p < .001$), and between pleasantness and gamma power in the temporal ROI ($r_{rm} = .477$, 95% CI [.26, .65], $p < .001$). In the temporal ROI, higher arousal ratings coincided with higher relative alpha power and lower relative gamma power. Each correlation was based on 15 participants and 77 paired observations.

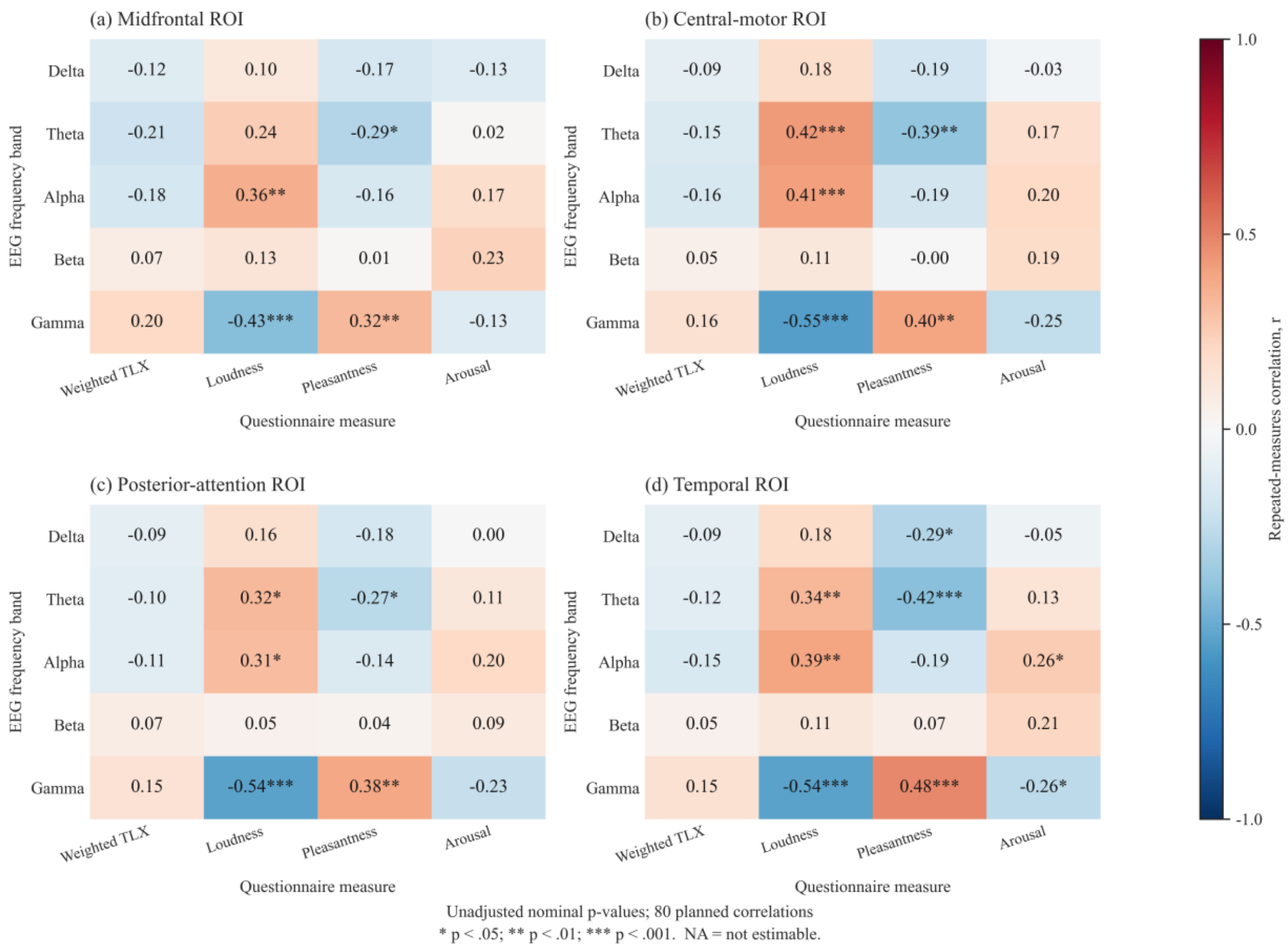

Fig. 13. Repeated-measures correlations ($r_{rm}$) between the questionnaire measures and relative EEG band power in the four ROIs across the six noise conditions. Asterisks indicate unadjusted $p$ values: * $p < .05$, ** $p < .01$, *** $p < .001$.

### 3.4.2 Associations between cognitive performance and EEG

Fig. 14 presents the correlations between the PI of each task and the relative band power recorded during the same task, so that each performance measure was paired with the EEG activity that accompanied it. Only two of the 80 correlations were significant, fewer than the four expected by chance: higher PI coincided with lower central-motor alpha power in Task 4 ($r_{rm} = -.264$, $p = .041$; 15 participants, 74 paired observations) and with lower central-motor gamma power in Task 3 ($r_{rm} = -.262$, $p = .049$; 14 participants, 70 paired observations). The two associations occurred in different tasks and bands, and all other coefficients were weak ($|r_{rm}| \leq .25$).

Although most correlations were not significant, several bands showed coefficients of the same sign across the four ROIs within a task, and these signs differed between tasks. In Task 1 (focused attention), PI was positively correlated with relative alpha power ($r_{rm} = .05$ to $.19$) and negatively with delta power ($r_{rm} = -.04$ to $-.17$), whereas in Task 4 (information extraction), PI was negatively correlated with alpha and beta power ($r_{rm} = -.13$ to $-.26$) and positively with delta power ($r_{rm} = .13$ to $.25$). In Task 2 (divided attention) and Task 3 (decision making), PI was negatively correlated with beta power ($r_{rm} = -.09$ to $-.24$) and gamma power ($r_{rm} = -.11$ to $-.26$), respectively. Because nearly all of these coefficients were non-significant, these tendencies are reported only descriptively.

Taken together, no consistent within-participant association between cognitive performance and EEG spectral power was found, in contrast to the numerous associations between EEG and the loudness and pleasantness ratings reported in Section 3.4.1. Because the strengths of the two sets of associations were not compared directly, however, this contrast does not show that EEG is more closely coupled to perceived responses than to performance.

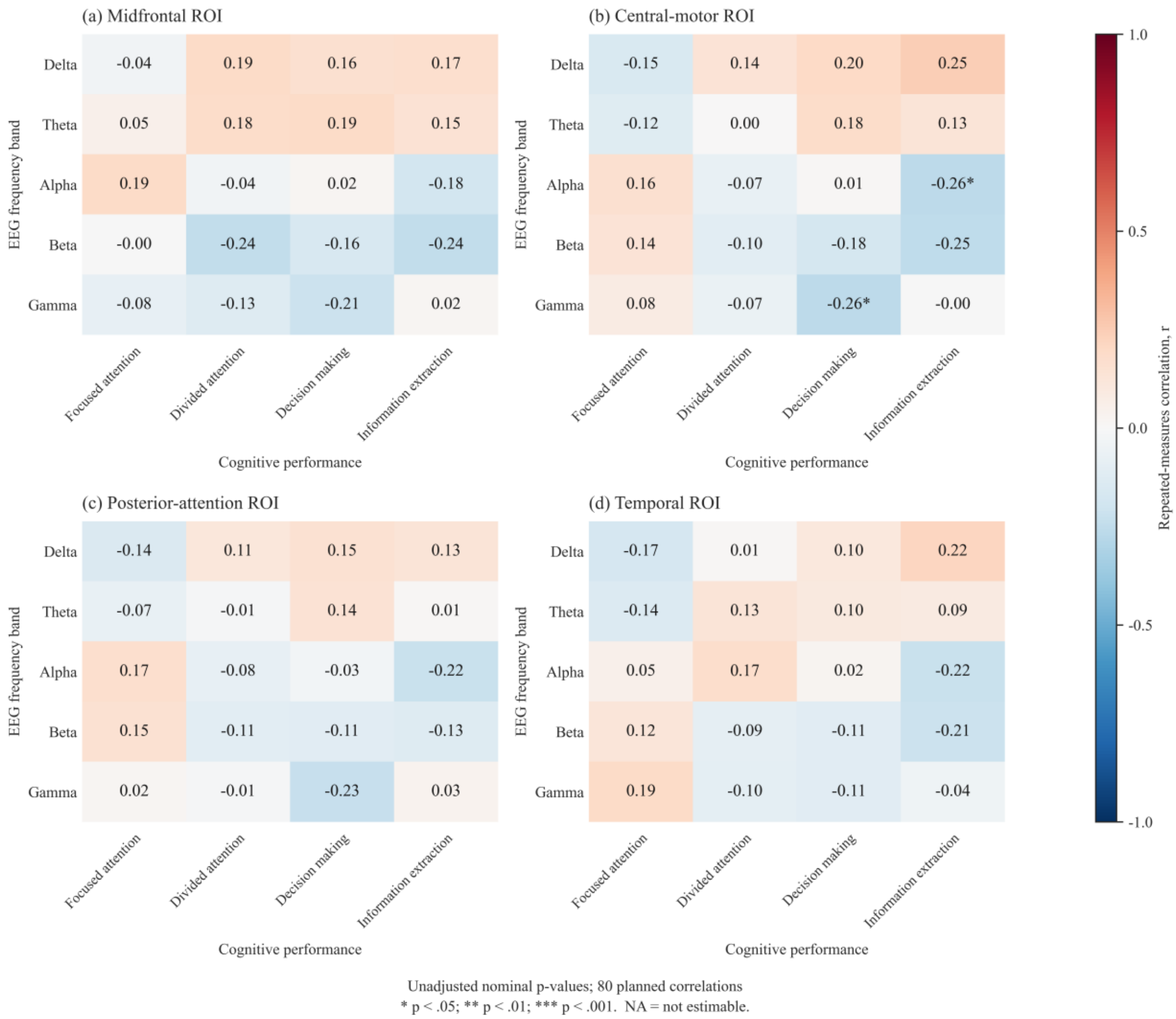

Fig. 14. Repeated-measures correlations ($r_{\mathrm{rm}}$) between the performance index of each task and the relative EEG band power recorded during the same task in the four ROIs across the six noise conditions. Asterisks indicate unadjusted $p$ values: * $p < .05$.

## 4. Discussion

This study measured perceived, cognitive, and EEG responses in the same participants under identical acoustic conditions to address three research questions: how each response varied with AC and RT noise at different levels (RQ1), whether the responses were consistent within participants (RQ2), and whether perceived and EEG responses could distinguish conditions that cognitive performance did not (RQ3). Performance did not differ significantly between conditions, whereas acoustic appraisal clearly did, perceived workload was higher under noise than under control, and task-period EEG differed between sources or levels in specific tasks and scalp regions. Within participants, EEG covaried mainly with loudness and pleasantness, rarely with arousal and performance, and not with workload. The following sections discuss each research question in turn and relate the findings.

## 4.1 Effects of noise source and level within each domain (RQ1)

Acoustic appraisal responded most clearly to the acoustic conditions. The noise conditions were rated louder, less pleasant, and more arousing than control, and all three ratings varied with sound level. This agrees with classroom studies in which students' soundscape evaluations distinguished between sound conditions (Pellegatti et al., 2026) and in which traffic noise affected subjective assessments (Chen & Ou, 2021). Differences between the two sources, however, were limited to loudness, which showed both a source effect and a source-by-level interaction. (Pellegatti et al., 2026), by contrast, found mechanical ventilation sound to be perceived as monotonous and road traffic as chaotic. The attributes retained did not include the monotonous and chaotic descriptors that separated the sources in that study, and the ratings were given after a block of four tasks, so they may have reflected the task context as well as the sound. Within the tested range, appraisal was therefore associated more with sound level than with source. Perceived workload showed a weaker pattern. Raw and weighted NASA-TLX scores were about 10 points higher under noise than under control, but overall workload did not vary with source or level; among the dimensions, only mental demand differed between sources, and physical demand showed a source-by-level interaction. This increase, together with the higher effort and temporal-demand ratings, is consistent with performance being maintained through additional effort (Hockey, 2011). Because the NASA-TLX was completed once per condition block, after all four tasks, each rating may also have averaged over demands that differed between tasks.

Cognitive performance did not differ significantly between noise and control, or between sources and levels, in any task, whether it was assessed with the performance index, accuracy, or response time. Impairments have been reported for children at comparable or higher levels (Caviola et al., 2021; Ljung et al., 2009; Papanikolaou et al., 2015), whereas (Pellegatti et al., 2026) found no accuracy effects at 47 dBA. Because the present levels overlapped those at which impairments were found, the difference is more likely to lie in the participants and tasks than in the sound level. The participants were university students, and children are more susceptible than adults to noise-induced disruption (Klatte et al., 2013). The tasks were also short and were repeated across seven conditions, which may have left spare capacity for performance to be protected. Studies using other outcomes have found effects of AC and traffic noise on visual-memory performance (Meng et al., 2021) and listening comprehension (Chen & Ou, 2021), so the present results apply to the four tasks tested rather than to learning in general.

Apart from loudness and two NASA-TLX dimensions, task-period EEG was the only response that differed between AC and RT. In Task 1, the AC–RT difference in posterior-attention power reversed direction across levels, and in Task 4, the midfrontal response was larger under RT than under AC. These results extend earlier task-period EEG work under air-conditioning and talking noise (Tsujimura & Ueno, 2010). Because power was normalized to control, the models compared the noise conditions with one another rather than with control. The EEG results therefore show that task-period power was sensitive to source, and less robustly to level, in some task and ROI combinations, not that either source has a consistent neural signature or that one task or region was more affected than another.

## 4.2 Consistency across response domains (RQ2)

The within-participant analyses indicate that the three types of response were only partly consistent. EEG was consistent with acoustic appraisal: across the six noise conditions, higher loudness and lower pleasantness ratings were accompanied by higher relative theta power and lower relative gamma power, and loudness was also positively associated with relative alpha power. By contrast, associations with arousal and same-task performance were rare, and none was found for perceived workload. At the condition level, appraisal and performance were also inconsistent, because appraisal changed with the acoustic conditions while performance did not. This pattern may partly reflect how much each response varied within participants. Among the noise conditions, loudness and pleasantness showed the largest level effects, arousal a smaller one, and workload and performance no significant variation. A response that changes little across conditions offers little within-participant variance to relate to EEG, so the scarcity of associations for workload, arousal, and performance does not show that these responses are unrelated to neural activity. Conversely, because association strengths were not compared directly, the results do not show that EEG is more closely linked to perception than to workload or cognition. The correlations were also not adjusted for source or level and may partly reflect a shared response to sound level.

The spectral pattern also needs careful interpretation. Because relative power expresses each band as a fraction of total power, higher low-frequency fractions and a lower gamma fraction may reflect a single shift in spectral balance, which can arise from aperiodic as well as oscillatory activity (Donoghue et al., 2020). Scalp gamma activity is also susceptible to muscle artifacts (Muthukumaraswamy, 2013). Moreover, theta has been linked to working-memory load and mental workload, particularly at frontal sites (Borghini et al., 2014; Klimesch, 1999), but here the theta associations appeared over central and temporal sites and involved loudness and pleasantness rather than rated workload. This supports the caution against assigning a single cognitive function to a frequency band (Chikhi et al., 2022). Previous studies have related subjective ratings to EEG during listening without a task (N. Zhang et al., 2025), an EEG ratio to attention and NASA-TLX scores (Mohammadi et al., 2023) , and lecture-period EEG to later examination results (Eqlimi et al., 2020). (Pellegatti et al., 2026) collected soundscape ratings and performance from the same students but left their combined analysis to future research. The present study provides such an analysis within participants, pairing task-period EEG with the ratings for the same condition and with performance on the same task. Differences in activity state and outcome mean that the directions observed here should not be treated as replications of those studies.

### 4.3 Differences beyond cognitive performance (RQ3)

The answer to RQ3 differs between responses. Acoustic appraisal distinguished noise from control and varied with sound level, although performance differed in neither comparison. Perceived workload also distinguished noise from control, but not the sources or levels. EEG distinguished some sources and levels in specific tasks and regions, again without corresponding differences in performance. However, these responses did not fully agree on which conditions differed: appraisal separated levels but, apart from loudness, not sources; workload separated only noise from control; and EEG separated sources and, less robustly, levels in some task and ROI combinations. Within participants, however, EEG covaried with the two ratings that varied most with level.

The Introduction outlined several mechanisms through which performance can remain stable while other responses change. The present results are compatible with some of them, although the design did not test any of them directly. The increase in workload, including higher effort ratings, is in the direction expected if performance was protected by additional effort (Hockey, 2011) , and stable performance is also compatible with attention being shielded against distraction (Halin et al., 2014) . Rated arousal bears on the account that (Pellegatti et al., 2026)  used to interpret their response-time effects, which rests on the inverted-U relation between arousal and performance (Yerkes & Dodson, 1908); they could not test it because they measured no indicator of arousal. Near the optimum of this relation, a sound may change arousal without changing performance. Here, rated arousal was higher under noise than under control and varied with level while performance remained unchanged, which is consistent with that account. EEG spectral power, however, covaried with rated arousal in only 2 of 20 tests, so the psychological and physiological measures converged only weakly. Because EEG spectral measures are not a direct index of arousal, peripheral indicators such as skin conductance (Alvar & Francis, 2024) may be better suited to testing this account.

These results show which responses distinguished conditions that performance did not; they do not rank the sensitivity of the response domains, which were measured on different scales, over different time windows, and with different numbers of tests. Nor do they show that performance was unaffected, because a non-significant difference is not evidence of equivalence (Lakens, 2017). What they do show is that conditions not distinguished by performance were nonetheless distinguished by how they were perceived. A non-significant performance result should therefore not be taken to mean that the acoustic conditions were equivalent for the students exposed to them.

### 4.4 Implications for assessing environmental noise in classrooms

Two implications follow for how the effects of environmental noise in classrooms are assessed. First, performance tests alone may miss differences in how students experience the acoustic environment. Here, conditions that produced no detectable difference in short cognitive tasks were clearly distinguished by acoustic appraisal, which matters for acoustic comfort in its own right (Pellegatti et al., 2023; Visentin et al., 2023). Perceived workload was also higher under noise than under control, suggesting that these conditions demanded more effort even though performance was maintained. Such assessments should therefore include perceived responses alongside performance. Second, EEG added information that the other measures did not, including differences between sources, but its sensitivity to analytical choices means that it is better treated as a research tool for studying how noise affects task processing than as a criterion for judging acoustic quality.

The comparison of AC and RT noise was motivated by the link between each source and the way a classroom is ventilated. Within the tested range, appraisal depended more on sound level than on source, suggesting that the level of sound reaching the classroom may matter more for perceived quality than whether it comes from an air-conditioning system or from traffic entering through open windows. This inference should be treated as a hypothesis. The experiment reproduced recordings through a loudspeaker with the windows closed and did not change ventilation mode, air quality, or thermal conditions, and air quality alone has been associated with student performance (Wargocki et al., 2020).

### 4.5 Limitations and future research

Several limitations qualify these findings. The sample was small: 16 university students were participated and contributed paired data to the cross-domain correlations, which limits the precision of all estimates and the power to detect small effects, particularly for workload and performance, and makes the EEG and cross-domain analyses exploratory. One classroom, one loudspeaker position, selected recordings of each source, and brief computer-based tasks limit generalization to other sources, rooms, and activities, and to children.

In addition, ROI labels describe scalp locations rather than neural sources, and the relative power used in the correlations and the control-normalized change used in the mixed models are different quantities, so the two sets of EEG results do not describe the same effect. Repeated-measures correlation also assumes a common within-participant slope (Bakdash & Marusich, 2017), and the questionnaires referred to a whole condition block, whereas EEG and performance referred to individual tasks. Future analyses could use multilevel models that separate within-condition from between-condition covariance, compare association strengths directly, separate periodic from aperiodic spectral components (Donoghue et al., 2020) , and verify artifact handling in the gamma range.

Finally, testing the mechanisms discussed in Section 4.3 would require additional measures. Eventfulness ratings, which capture the second main dimension of soundscape perception (Axelsson et al., 2010), would show whether the distinction between monotonous and chaotic sounds reported in classrooms (Pellegatti et al., 2026) also applies to AC and RT noise, and peripheral indicators of arousal would allow the arousal account to be tested directly. Field studies in which the ventilation mode is actually changed would show whether the present contrasts hold when sound, air quality, and thermal conditions change together.

## 5. Conclusions

This study examined whether perceived, cognitive, and EEG responses to AC and RT noise are consistent within the same participants, and whether perceived and EEG responses can distinguish acoustic conditions that cognitive performance does not. Sixteen university students completed four cognitive tasks in a classroom under a no-noise control condition and under AC and RT noise, each played back at 55, 60, and 65 dBA. Cognitive performance did not differ significantly between the acoustic conditions in any task. By contrast, the noise conditions were rated louder, less pleasant, and more arousing than the control condition, and these ratings depended more on sound level than on noise source. Perceived workload was also higher under noise than under control, although it did not vary with noise source or level. Task-period EEG power differed between noise sources and levels in specific tasks and scalp regions; the source effects were largely robust to the different treatment, whereas the level effects were not.

The three types of response were only partly consistent. Within participants, relative EEG band power covaried mainly with loudness and pleasantness, whereas associations with arousal and cognitive performance were rare and none were found for perceived workload. Overall, acoustic conditions that cognitive performance did not distinguish were still distinguished by perceived responses and, in part, by EEG. A non-significant difference in performance should therefore not be taken to mean that the acoustic conditions are equivalent for the students exposed to them, and assessments of environmental noise in classrooms should include perceived responses alongside cognitive performance. These findings should be confirmed in larger samples; future work should also separate within-condition from between-condition covariance and test directly whether performance is equivalent across conditions.